\documentclass[useAMS,usenatbib]{mn2e}
\usepackage{txfonts}

\newcommand{\Msun}{$M_{\odot}$}

\usepackage{graphicx}
\usepackage{longtable}

\title[ALMA spectra of SN 1987A]
  {ALMA spectral survey of Supernova 1987A --- molecular inventory, chemistry, dynamics and explosive nucleosynthesis}
\author[Matsuura et al.]
{M. Matsuura$^{1,2}$, 
R. Indebetouw$^{3,4}$,
S. Woosley$^{5}$,
V. Bujarrabal$^6$,
F.J.  Abell{\' a}n$^7$,
R. McCray$^{8}$,
             \newauthor 
J. Kamenetzky$^{9}$,
C. Fransson$^{10}$,
M. J. Barlow$^2$,
H.L. Gomez$^1$,
P. Cigan$^1$,
I De Looze$^2$, 
             \newauthor 
 J. Spyromilio$^{11}$,
L. Staveley-Smith$^{12, 13}$,
G. Zanardo$^{12}$, 
P. Roche$^{14}$,
J. Larsson$^{15}$,
             \newauthor 
S. Viti$^{2}$,
J.Th. van Loon$^{16}$,
J.C. Wheeler$^{17}$,
M. Baes$^{18}$, 
R. Chevalier$^3$,
P. Lundqvist$^{10}$,
            \newauthor 
J. M. Marcaide$^{7}$,
E. Dwek$^{19}$,
M. Meixner$^{20,21}$,
C.-Y. Ng$^{22}$,
G. Sonneborn$^{19}$,
J. Yates$^2$,
 \\
  $^1$ School of Physics and Astronomy, Cardiff University, Queen's Buildings, The Parade, Cardiff, CF24 3AA, UK \\
  $^2$ Department of Physics and Astronomy, University College London, Gower Street, London WC1E 6BT, UK \\
  $^3$ Department of Astronomy, University of Virginia, P.O. Box 400325, Charlottesville, VA 22904-4325, USA \\
  $^4$ National Radio Astronomy Observatory, 520 Edgemont Rd, Charlottesville, VA 22903, USA \\
  $^{5}$ Department of Astronomy and Astrophysics, University of California, Santa Cruz, CA 95064, USA \\
  $^6$ Observatorio Astron\'{o}mico Nacional (OAN-IGN), Apartado 112, E-28803 Alcal\'{a} de Henares, Spain \\
  $^7$ Universidad de Valencia, C/Dr. Moliner 50, E-46100 Burjassot, Spain \\
  $^{8}$ Department of Astronomy, University of California, Berkeley, CA 94720-3411, USA \\
  $^{9}$ Westminster College, Department of Physics, 1840 South 1300 East, Salt Lake City, UT 84105, USA \\
  $^{10}$ Department of Astronomy, The Oskar Klein Centre, Stockholm University, Alba Nova University Centre, SE-106 91 Stockholm, Sweden \\
  $^{11}$ European Southern Observatory (ESO), Karl-Schwarzschild-Strasse 2, 85748 Garching, Germany \\
  $^{12}$ International Centre for Radio Astronomy Research (ICRAR), The University of Western Australia, Crawley, WA 6009, Australia \\
  $^{13}$ ARC Centre of Excellence for All-Sky Astrophysics (CAASTRO), Australia \\
  $^{14}$ Department of Physics, University of Oxford, Oxford OX1 3RH, UK \\ 
  $^{15}$ KTH, Department of Physics, and the Oskar Klein Centre, AlbaNova, SE-106 91 Stockholm, Sweden \\
  $^{16}$ School of Physical and Geographical Sciences, Lennard-Jones Laboratories, Keele University, Staffordshire ST5 5BG, UK \\
  $^{17}$ Department of Astronomy, University of Texas, Austin, TX 78712-0259, USA \\
  $^{18}$ Sterrenkundig Observatorium, Universiteit Gent, Krijgslaan 281 S9, B-9000 Gent, Belgium \\
  $^{19}$ Observational Cosmology Lab, Code 665, NASA Goddard Space Flight Center, Greenbelt, MD 20771, USA \\
  $^{20}$ Space Telescope Science Institute, 3700 San Martin Dr., Baltimore, MD 21218, USA \\
  $^{21}$ Department of Physics and Astronomy, The Johns Hopkins University, 366 Bloomberg Center, 3400 N. Charles Street, Baltimore, MD 21218, USA \\
  $^{22}$ Department of Physics, The University of Hong Kong, Pokfulam Road, Hong Kong, China \\
}
\date{Released 2016 Xxxxx XX}

\pagerange{\pageref{firstpage}--\pageref{lastpage}} \pubyear{2016}

\def\LaTeX{L\kern-.36em\raise.3ex\hbox{a}\kern-.15em
    T\kern-.1667em\lower.7ex\hbox{E}\kern-.125emX}

\begin{document}

\label{firstpage}

\maketitle

\begin{abstract}
 We report the first molecular line survey of Supernova 1987A in the millimetre wavelength range.
 In the ALMA 210--300 and 340--360\,GHz spectra, we detected cold (20--170\,K) CO, $^{28}$SiO, HCO$^+$  and SO, with weaker lines of $^{29}$SiO from ejecta.
 This is the first identification of HCO$^+$ and SO in a young supernova remnant.
 We find a dip in the $J$=6--5 and 5--4 SiO line profiles, suggesting that the ejecta morphology is likely elongated.
The difference of the CO and SiO line profiles is consistent with hydrodynamic simulations, which show that Rayleigh-Taylor instabilities cause mixing of gas, 
 with heavier elements much more disturbed, making more elongated structure.
 We obtained isotopologue ratios of $^{28}$SiO/$^{29}$SiO$>13$, $^{28}$SiO/$^{30}$SiO$>14$, and  $^{12}$CO/$^{13}$CO$>$21,
with the most likely limits of $^{28}$SiO/$^{29}$SiO$>128$, $^{28}$SiO/$^{30}$SiO$>189$.
Low $^{29}$Si and $^{30}$Si abundances  in SN 1987A are consistent with nucleosynthesis models that show inefficient formation of neutron-rich isotopes in a low metallicity environment, such as the Large Magellanic Cloud. 
The deduced  large mass of HCO$^+$ ($\sim5\times10^{-6}$\,\Msun) and small SiS mass ($<6\times10^{-5}$\,\Msun) might be explained by some mixing of elements immediately after the explosion.
The mixing might have  caused some hydrogen from the envelope to sink into  carbon and oxygen-rich zones after the explosion, enabling the formation of a substantial mass of HCO$^+$. Oxygen atoms may have penetrated into silicon and sulphur zones, suppressing formation of SiS.
Our ALMA observations open up a new window to investigate chemistry, dynamics and explosive-nucleosynthesis in supernovae.
\end{abstract}


\section{Introduction}

Core-collapse supernovae (SNe) play a key role in the evolution of the interstellar medium (ISM) of galaxies.
First, heavy elements are synthesised in the stellar interior, and newly synthesised elements are expelled from SNe, enriching the ISM with metals and dust
\cite[e.g.][]{Kobayashi:2011hja, Dwek:1998p3931}.
Second, the interaction of the fast expanding ejecta into the ambient ISM results in shocks \citep[e.g.][]{Gotthelf:2001bn}.
However, there are still  limited studies of how turbulent gas evolves in supernova remnants before interacting with a circumstellar medium and subsequently the ISM, or measurements of the elements that have been synthesised in individual SNe.

So far, most abundance measurements of SN remnants have focused on detecting lines at X-ray, UV and optical wavelengths to derive elemental abundances in SN remnants \citep[e.g.][]{Reynolds:2008fi, Grefenstette:2014ds}.
Constraints obtained from these short wavelength spectra  are limited to the atomic lines of the main isotope only. 
Recently, our ALMA Cycle-0 programme discovered cold $^{28}$SiO and $^{12}$CO, and possibly  $^{29}$SiO emission from the ejecta of SN 1987A \citep{Kamenetzky:2013fv}.
Because the  line shift of an isotopologue can be larger than the SN expansion velocity  only at millimetre and submillimetre wavelengths,
ALMA provides a unique opportunity to investigate isotope abundances in SNe.

The formation of molecules in SNe was reported first in  SN\,1987A \citep[e.g.][]{Spyromilio:1988hk, Roche:1989jb, Lepp:1990cz}, followed by  a handful of detections in other supernovae within a few years after the explosion.
Near- and mid-infrared observations detected only small masses of molecules \citep[$\sim10^{-3}$\,\Msun; ][]{Lepp:1990cz, Kotak:2006p13473} in a few young ($<$2 yr) SNe.  
Recently, a large mass ($>$0.01\,\Msun) of cold ($<$120\,K) CO molecules was reported in the ejecta of SN 1987A 25 years after the explosion \citep{Kamenetzky:2013fv}.
Also a large mass ($\sim$0.5\,\Msun) of cold ($\sim$20\,K) dust \citep{Matsuura:2011ij, Indebetouw:2013vv} was recorded in the ejecta 23--25 years after the explosion.
These findings show that the SN ejecta have entered a cold molecular phase, which can be investigated with ALMA.
Additionally, vibrationally excited H$_2$ emission was detected recently, showing the presence of H$_2$ in the ejecta \citep{Fransson:2016jb}.
These later-phase observations can provide insights into molecular chemistry in the ejecta, that can be compared with chemical models
 \citep[e.g.][]{Lepp:1990cz, Rawlings:1990th, Cherchneff:2008gl, Cherchneff:2009ex, Clayton:2011fp, Sarangi:2013bj}.

It is believed that hydrodynamical instabilities at the time of the supernova explosion and shortly thereafter disrupted the nuclear burning zones into clumps of different compositions \citep[e.g.][]{Hammer:2010di, Wongwathanarat:2015jv}.
Microscopic mixing occurs at the thin interfaces of regions of different nuclear burning zones.
It has also been suggested that macroscopic mixing, triggered by these instabilities \citep{Li:1993eq} has caused the different elemental zones to have similar velocity profiles \citep{McCray:1993p29839}.
After a few weeks, the instabilities should have died out with the clumps maintaining their initial motion up to the present.
The CO expansion velocities on days$~$100 and 10,000 after the explosion were measured to be the same 
\citep{Liu:1992p29944, Kamenetzky:2013fv}.\
\citet{Chevalier:1979kp} and 
\citet{Milisavljevic:2015fw} have suggested that such clumps with discrete elemental abundance can be sustained even in the 450 years-old Galactic supernova remnant, Cassiopeia A.
Our ALMA measurements of velocities from different molecules can test whether mixing has caused the different elemental zones to have similar velocity profiles.

An outstanding question is the extent to which microscopic and macroscopic mixings occurs immediately after the explosion.
Recent 3-D explosion modelling has proposed that Rayleigh-Taylor instabilities of the gas can cause some mixing of gas between zones at the time of the explosion \citep{Fryxell:1991p29866, Hammer:2010di}. 
Chemical models predict that microscopic mixing could result in different compositions of molecules, allowing the formation of hydrides such as OH, OH$^+$ and CH, and opening up  new chemical pathways to increase the CO mass  \citep{Lepp:1990cz}.
If these hydrides are present, their line emission could be detected in an unbiased spectral line survey, thus testing the early evolution of the ejecta triggered by instabilities in the SN explosions.

SN1987A is  unique. It is located only 50\,kpc away.
 Its inner ejecta has not yet mixed with the circumstellar material, expelled by its progenitor before the explosion.
Given the early stage of the remnant evolution, the blast wave has just overtaken the high density circumstellar material in the equatorial ring
 \citep{France:2010p29071, Ng:2013bt, Fransson:2015gp}.
 CO and SiO detected with ALMA's spectral-imaging \citep{Kamenetzky:2013fv}
have unambiguously formed in the ejecta rather than in the ISM or circumstellar material swept-up by  the SN remnant. 
The majority of SN\,1987A's ejecta has not yet experienced a reverse shock \citep{France:2010p29071}.
After passage through reverse shocks, many molecules in the ejecta may be destroyed, 
though some molecules may reform after passage through reverse shocks, 
as found in Cassiopeia A \citep{Rho:2012br, Wallstrom:2013he, Biscaro:2014kh}.

We present a molecular line survey of SN\,1987A's ejecta in the 210--300 and 340--360\,GHz ALMA spectra,
from Cycle-1, Cycle-2 and Cycle-3 observations.
The plan of paper is  (a) mass of molecules, (b) isotope ratio and (c) line profiles, followed by 
discussions of molecular chemistry,  line profile, and isotope ratio.

\section{Observations and data reduction}

ALMA obtained spectra covering a continuous spectral range from 210 to 300\,GHz,
using a combination of 10 ALMA `Scheduling Blocks'  in bands 6 and 7. 
 As part of the ALMA cycle 2 program, 2013.1.00280.S (P.I. Matsuura),
 the spectra of SN\,1987A were obtained
on 2014 August 29th (day 10,048 since the explosion) for part of band 7 (275--298\,GHz), and on 2014 September 2nd (day 10,053) for  band 6
(211--275\,GHz).
Additionally, 230\,GHz CO J=2--1 data were obtained on 2014 August 18th (day 10,037), as part of the ALMA cycle 1 programme
2012.1.00075.S  (P.I.  Indebetouw), where the spectrum covering 	CO $J$=3--2 and 357\,GHz HCO$^+$ J=4--3
was obtained on 2015 July 25th (day 10,378) as part of the ALMA cycle 3 programme 2013.1.00063.S (P.I.  Indebetouw).
 A series of slightly overlapping $\sim$2000 km\,s$^{-1}$ wide segments were observed in sequence;
whereas a typical SN\,1987A ejecta line width is 2000 km\,s$^{-1}$ \citep[Full width of half maximum (FWHM); e.g. ][]{Kamenetzky:2013fv}, 
so that one emission line can fill one ALMA spectrum segment.

We requested a root mean square (rms) noise level of  47--52\,$\mu$Jy, which was in general achieved. 
The integration time was  480--750\,s per segment,
and the total observing time was 30--34\,min for band 6 segments and 40\,min for band 7 segments, with a total observing time of 5.6\,hours.
The combined spectrum is the result of 10$\times$4 of individual settings and observations for bands 6 and 7 with ALMA.
Each setting and observation was reduced individually with the Common Astronomy Software Applications ({\sc CASA}) in the 
post observation process,
before being combined into a single spectrum.
The baselines were weighted in order to have a fixed circular beam size (0.3'') across the frequencies.
That is slightly smaller than the naturally weighted beams, which have typically 0.3--0.5\,arcsec elliptical beams.
The spectra have been smoothed to enhance  the signal-to-noise ratio, yielding a final spectral resolution of 345.2\,km\,s$^{-1}$.

\begin{figure*}
\centering
\resizebox{\hsize}{!}{\includegraphics{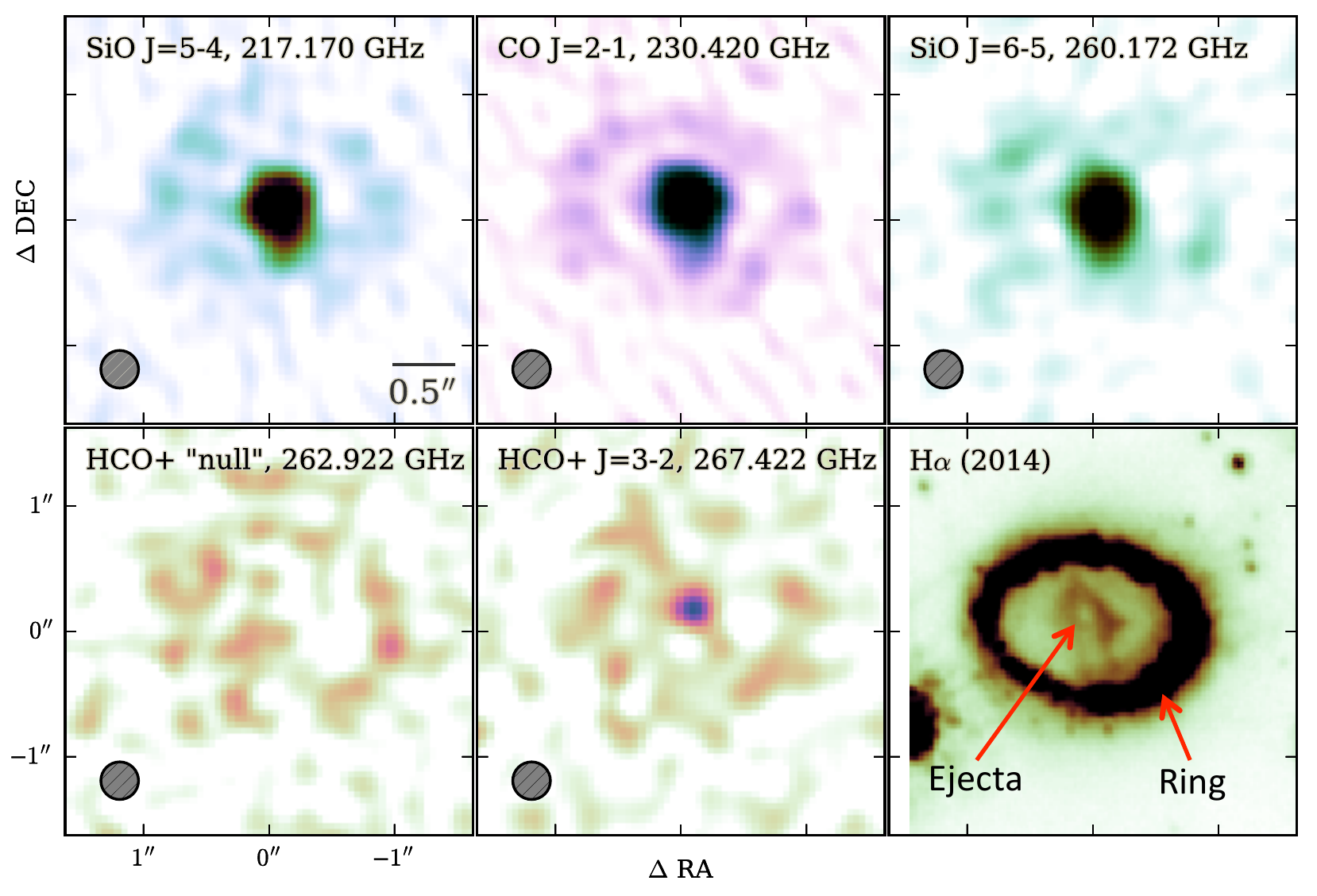}}
\caption{ The ALMA 217, 230 and 267\,GHz images, along with a HST H$\alpha$ image \citep{Fransson:2015gp, Larsson:2016wg}.
The SiO and CO lines originate from the ejecta, located in the centre of the ring.
The faint emission from the ring seen in all ALMA images, is due to synchrotron radiation from the ring.
The quoted frequencies are those observed, rather than the line centre.
The HCO$^+$ 267.4\,GHz line is clearly detected in the ejecta, when comparing with the  `null' frequency image at 262.9\,GHz.
\label{fig-ch}}
\end{figure*}

SN 1987A is spatially resolved, with a clear distinction between the ejecta
 and the ring (Fig.\, \ref{fig-ch}).
The spectrum was extracted from 
an ellipse-shaped aperture with long axis of  1.29\,arcsec (east-west) and short axis of 1.10\,arcsec (north-south),
centred on the SN ejecta, thereby minimising the contamination by emission from the ring.
However,
there is some slight leak of the ring flux into the aperture, due to imperfect image reconstruction,
contributing some continuum level to the 'ejecta' spectrum.
The largest contribution to the noise is the random level of this ring flux `leak' into the aperture.
We evaluated uncertainties by calculating the $\sigma$ level by taking the difference
of fluxes at four different apertures along the ring, and found that this noise level is on average about 80\,\%
larger than the rms noise alone. We adopt this aperture measured noise level as the noise in the analysis.

\section{Spectrum and images}

\subsection{ALMA spectrum and molecular identifications}

\begin{figure*}
\centering
\resizebox{\hsize}{!}{\includegraphics{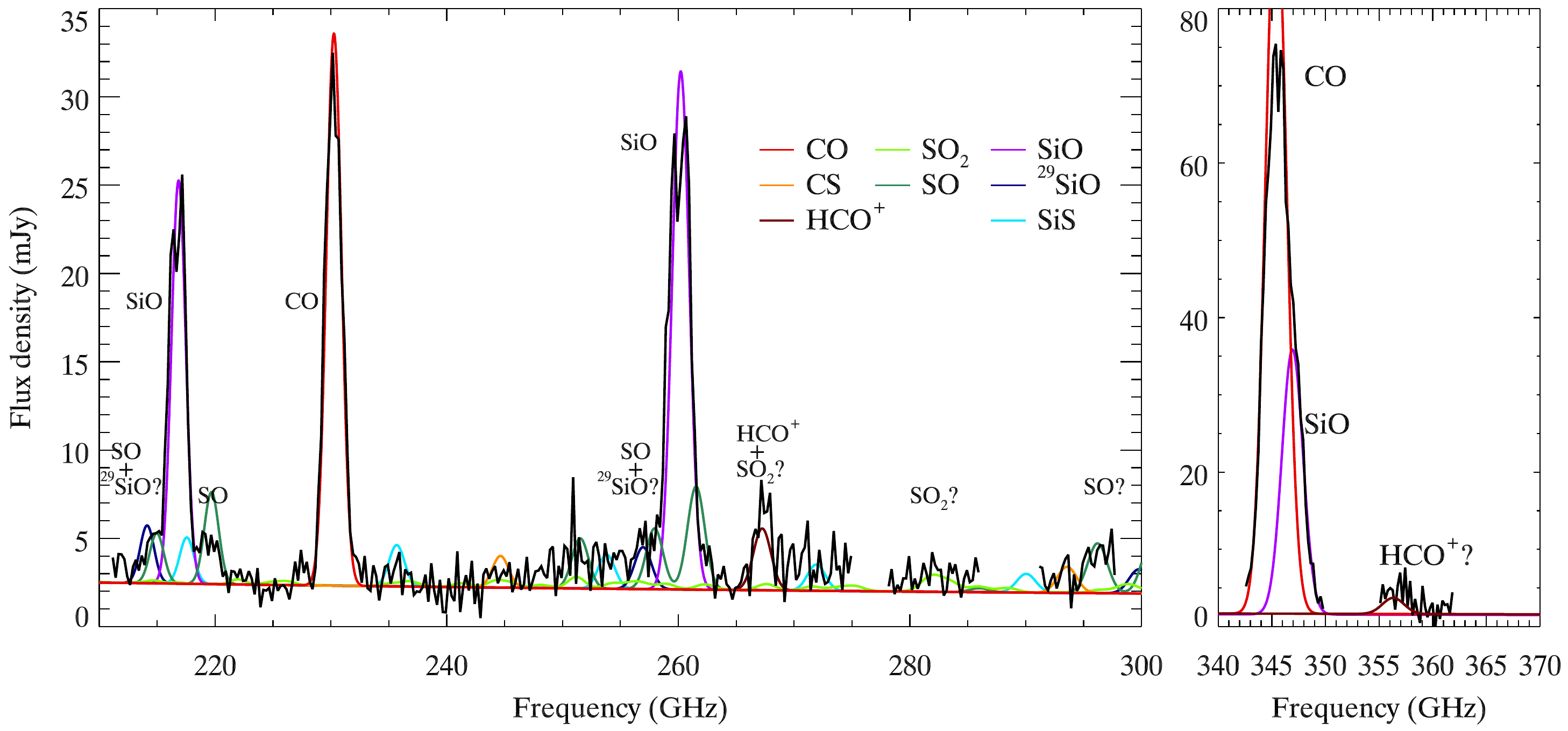}}
\caption{ The ALMA 210--300 and 340-360\,GHz spectra of SN\,1987A's ejecta (black).
Major molecular features are labeled. Model molecular spectra with a 2000\,km\,s$^{-1}$
FWHM Gaussian line profile  and the excitation temperature of 40\,K, are plotted in colour, guiding molecular identifications. Underlying synchrotron radiation,
leaked from the ring into the aperture, is represented by a power low with an index of $-0.8$.
\label{fig-spec}}
\end{figure*}
\begin{table*}
  \caption{   The lines identified in SN 1987A, together with some representative non-detected lines \label{table-flux}}
\begin{center}
 \begin{tabular}{l  r r{@}{$\pm$}l{@}{$\pm$}l r{@}{$\pm$}l r r r r r}
\hline
Line ID 
& $\nu_0$      
& \multicolumn{3}{c}{Int.} 
& \multicolumn{2}{c}{$h$} 
& \multicolumn{1}{c}{$v_c$} 
& FWHM
\\ 
& GHz
& \multicolumn{3}{c}{$10^{-20}$W\,m$^{-2}$} 
& \multicolumn{2}{c}{mJy}
& km\,s$^{-1}$
& km\,s$^{-1}$
\\ \hline
SO $J_K$=$5_5$--$4_4$        & 215.221  &  2.6&0.3&0.2   \\
+ $^{29}$SiO $J$=5--4?       & 214.386     \\
SiO $J$=5--4                 & 217.105  & 35.1&0.4&0.2             & 28.7&0.5 & 220  & 1770 \\
SO $J_K$=$5_6$--$4_5$        & 219.949  &  3.3&0.4&0.3   \\
CO $J$=2--1                  & 230.538  & 51.0&0.4&0.3             & 30.0&0.1 & 130  & 2130 \\
SO $J_K$=$6_6$--$5_5$        & 258.256  & 1.8&0.6&0.0  \\
+ $^{29}$SiO $J$=6--5 ?      & 257.255  \\
SiO $J$=6--5                 & 260.518  & 54.1&0.6&0.0             & 29.1&0.4 & 230 & 2210 \\
HCO$^+$  $J$=3--2               & 267.558  & 6.7&0.6&0.1              &  3.8&0.3 & $-$200 & 2210 \\
+SO$_2$ $J_{K_{-}},_{K_{+}}$=13$_{3,11}$--13$_{2,14}$?        
                             & 267.537  & \multicolumn{3}{c}{} & \multicolumn{2}{c}{}  &  150  \\
CO $J$=3--2                  & 345.796  & 178&2.6&--- & 73&1.2 \\
+ SiO $J$=8--7               & 347.331   \\  \hline
{\it Upper limits of non-detections} \\
$^{13}$CO $J$=2--1           & 220.399 & \multicolumn{2}{c}{$<2.4$}\\
$^{30}$SiO $J$=5--4          & 211.853 & \multicolumn{2}{c}{$<2.5$} \\
$^{30}$SiO $J$=6--5          & 254.216 & \multicolumn{2}{c}{$<6.2$} \\
$^{30}$SiO $J$=7--6          & 296.575 & \multicolumn{2}{c}{$<3.3$} \\
CS $J=$5--4	                 & 244.936 & \multicolumn{2}{c}{$<3.1$} \\
CS $J=$6--5                  & 293.912 & \multicolumn{2}{c}{$<6.1$} \\
SO $J_K$=$6_5$--$5_4$        & 251.826 & \multicolumn{2}{c}{} \\
SO $J_K$=$5_4$--$4_5$ ?      & 294.768  & \multicolumn{2}{c}{} \\
SO$_2$ $J_{K_{-}},_{K_{+}}$=16$_{0,16}$--15$_{1,15}$  & 283.465 & \multicolumn{2}{c}{$<3.8$} \\
SiS $J=$13--12                   & 235.961 & \multicolumn{2}{c}{$<2.5$} \\
SiS $J=$14--13                   & 254.103 & \multicolumn{2}{c}{$<4.4$} \\
SiS $J=$15--14                   & 272.243 & \multicolumn{2}{c}{$<5.4$} \\
{\it HCO$^+$ $J$=4--3}       & 356.734  \\ 
\hline
\end{tabular}\\
\end{center}
$\nu_0$: the frequency of the molecular lines in vacuum.
Int.: the integrated line intensities. The first flux uncertainties are estimated from 
an Monte Carlo method of placing multiple-apertures on the images, and the second uncertainties are systematic errors
due to uncertainties of baseline determinations.
For the measurements of SO and $^{29}$SiO lines, neighbouring SiO lines were fitted with Gaussian profiles, and subtracted
from the spectra, before measuring the integrated line fluxes of SO, $^{29}$SiO lines.
$h$, $v_c$ and FWHM: the height, the central velocity and the FWHM of the Gaussian fits to the line profiles.
The LMC systemic velocity of 275\,km\,s$^{-1}$ was subtracted to obtain the central velocities.
The spectral resolution was 345.2\,km\,s$^{-1}$, and the uncertainties of these velocities are $\sim$100\,km\,s$^{-1}$.
For $^{28}$SiO, the dips are masked by the fitting, so that the measured Gaussian heights are higher than the peaks
observed in the spectra (Fig.\ref{fig-spec}).
Upper limits (3-$\sigma$) are given for non-detections,
assuming  Gaussian line profiles with a FWHM of 2000\,km\,s$^{-1}$.
\normalsize
\end{table*}

In figure\,\ref{fig-spec} we show the observed spectrum of SN\,1987A's ejecta, showing a variety of 
broad (FWHM$\sim$2000\,km\,s$^{-1}$) molecular lines.
The strongest molecular emissions are attributed to SiO and CO.
The emission at 267\,GHz is associated with HCO$^+$ $J$=3--2, with some possible contamination from SO$_2$.
Other weak features are due to SO and $^{29}$SiO. SO$_2$ and CS may contribute
some weaker features, though as discussed later that  is very unlikely.
The line identifications are summarised in Table\,\ref{table-flux}.

In order to identify molecular lines in our spectra,
in addition to simple matching of the frequencies of species to features present,
we have used molecular spectra predicted by the  non-LTE (local thermal equilibrium)  code {\sc RADEX} \citep{vanderTak:2007be}, 
including CO, SiO, $^{29}$SiO, SO, HCO$^+$, SiS, CS and SO$_2$
and the LTE code described by \citet{Matsuura:2002p25132} for SO$_2$.
The details of {\sc RADEX} modelling are described in Section\,\ref{analysis}.
We used  the {\sc HITRAN} molecular line list  for SO$_2$ modelling \citep{Rothman:2009p28626}.
The  temperatures of the models are 40\,K in this plot.

The comparison of the ALMA spectra with the model spectra indicates
that the weak features are indeed molecular lines.
The features at 215, 219, 258\,GHz and potentially at  252 and 261\,GHz,  are associated with SO.
Because the model predicts that the 215\,GHz line intensity cannot be explained only with
SO, an additional contribution from $^{29}$SiO is needed,
as suggested by \citet{Kamenetzky:2013fv}.
A similar blending issue of SO and $^{29}$SiO is also found for the 258\,GHz feature.

Another SiO isotope, $^{30}$SiO, has three transitions within the observed ALMA spectral coverage;
at 211.9\,GHz ($J$=5--4), 254.2\,GHz ($J$=6--5) and 296.6\,GHz ($J$=7--6). 
These lines are not detected, and upper limits for the line intensities are
listed in  Table\,\ref{table-flux}.

 $^{13}$CO has a transition at 220.399\,GHz. Our spectra show a feature that is comparable with this.
However, the line centre is slightly offset to shorter frequency to be $^{13}$CO.
This feature can probably be attributed to SO $J,K$=(5,6)--(4,5) at 219.9494\,GHz.
 No other $^{13}$CO transitions occur in our current frequency coverage, 
 so observations at other frequencies are required  to rule out the presence of $^{13}$CO emission.

The 267\,GHz feature is associated with HCO$^+$ with a potential contribution from SO$_2$.
This line is the only HCO$^+$  feature present in our spectral coverage, with  a marginal detection of HCO$^+$  $J$=4--3 at 357\,GHz,
whereas SO$_2$ has multiple transitions.
We display the SO$_2$ model spectrum in Fig.\,\ref{fig-spec} for comparison with the ALMA observed spectrum (Fig.\ref{fig-spec}).
The model spectrum predicts that the strongest  SO$_2$ feature would appear
at about 282\,GHz, but in the ALMA spectrum, this feature is not  detected.
This suggests that although SO$_2$ emission lines can be present across the 200--300\,GHz range, 
the contribution of  SO$_2$  to the 267\,GHz line is  small.
The 267\,GHz line is attributed mainly to HCO$^+$.

Early in its evolution (from day 377 up to 574), tentative detections of near-infrared CS vibrational-rotational transition 
were reported for SN\,1987A \citep{Meikle:1989vx, Meikle:1993tz}.
CS molecules have rotational transitions  at 244.9\,GHz and 293.9\,GHz,
but these lines were not detected in our ALMA spectrum.

We measured integrated line intensities after local continuum subtraction, as listed in Table\,\ref{table-flux}.
We fit the lines with a Gaussian, and their heights, centres and FWHMs are also listed in the table.
For non-detected lines, 3-$\sigma$ upper limits are given in the table.

\subsection{SiO and CO line profiles}

\begin{figure}
\centering
\resizebox{\hsize}{!}{\includegraphics{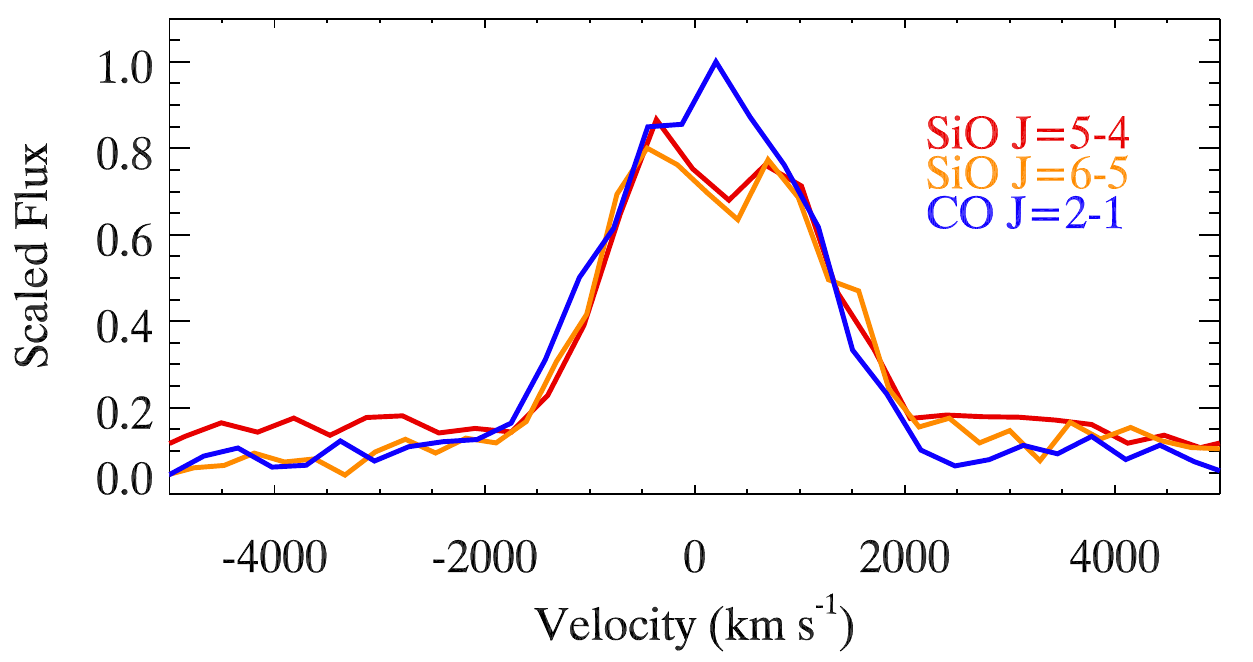}}
\caption{ The line profiles of SiO $J$=5--4 and 6--5, compared with that of CO $J$=2--1.
The SiO lines clearly show dips at the top at $\sim$300\,km\,s$^{-1}$.
An LMC systemic velocity of 275\,km\,s$^{-1}$ has been subtracted from all spectra.
\label{fig-line-profile}}
\end{figure}

The line profiles of  SiO $J$=5--4 and 6--5 are clearly  non-Gaussian, with dips at the top
(Fig.\,\ref{fig-spec}).
The enlarged SiO line profiles are displayed in Fig.\ref{fig-line-profile},
with a comparison with the CO line profiles. 
The profiles are plotted in velocity, after the LMC systemic velocity 
of 275\,km\,s$^{-1}$ \citep{vanderMarel:2002dj, Marshall:2004p9396, Evans:2015dn} was subtracted. 
That is a very small difference from the systemic velocity measured in the ring: 
286.5\,km\,s$^{-1}$ in Barycentre, i.e., 271\,km\,s$^{-1}$ in kinematic $v_{\rm LSR}$ with ALMA \citep{Groningsson:2008jba}.
However, compared with the ALMA velocity grid of 345.2\,km\,s$^{-1}$, this difference is negligible.
A dip at $\sim300$\,km\,s$^{-1}$ is prominent in the SiO line profiles, 
and the SiO $J$=5--4 and 6--5 line profiles resemble each other,
including the widths and depths of the dips.
The SiO dips are slightly offset to the red from line centre by $\sim$300\,km\,s$^{-1}$.

There is a hint of a dip at the shoulder of CO $J$=2--1 at $\sim-$100\,km\,s$^{-1}$.
Its shape changes with different re-sampling of the ALMA frequency, thus,
it is insufficient to verify whether this dip is real or not with CO $J$=2--1 alone.
A dip is also present in the CO $J$=3--2 spectrum (Fig.\ref{fig-spec}), and that corresponds to about $-$100\,km\,s$^{-1}$,
the same velocity as CO  $J$=2--1.
CO might have a plausible dip at about $\sim-$100\,km\,s$^{-1}$.

The FWHMs of the SiO lines were measured to be 1770 and 2210\,km\,s$^{-1}$ for J=5--4 and 6--5, respectively.
The difference  may not be real, because the tops of the line profiles were masked to remove the effect of the dip.
The CO FWHM velocity was measured to be 2130\,km\,s$^{-1}$. The SiO and CO FWHM are more or less consistent.

Figure\,\ref{fig-line-profile} shows that the axis of line symmetry is offset from the centre by 100--200 km\,s$^{-1}$
(also listed in Table\,\ref{table-flux}).
A similar offset is also seen in the 1.64-$\mu$m [Si\,{\small I}]+[Fe\,{\small II}] line profile 
 \citep{Kjr:2010p29878, Larsson:2016wg}.

\subsection{Images}

Figure\,\ref{fig-ch} shows the ALMA images at 217.255\,GHz (SiO $J$=5--4) and 
259.679\,GHz (SiO $J$=6--5), as well as at 230.615\,GHz (CO $J$=2--1).
These figures show clearly that the strong molecular emissions originate from the inner ejecta.
The ejecta are resolved spatially, much larger than the beam size, and appears to be slightly elongated to approximately south-north, following the elongation
direction of the H$\alpha$ image (right bottom of Fig.\,\ref{fig-ch}).
The 267.453\,GHz  image for the band covering HCO$^+$ $J$=3--2 can be compared with 
the 263.078\,GHz image, which does not have a contribution from the HCO$^+$ line. The difference of the contrast
between the ring and the ejecta in these two images shows that there are molecular lines
contributing in the 267.453\,GHz band, indicating the detection of HCO$^+$ $J$=3--2.
The faint emission seen in the ring is associated with synchrotron radiation
\citep[e.g.][]{Zanardo:2014gu}.

\section{Analysis of the line intensities} \label{analysis}

\begin{figure}
\centering
\resizebox{0.98\hsize}{!}{\includegraphics{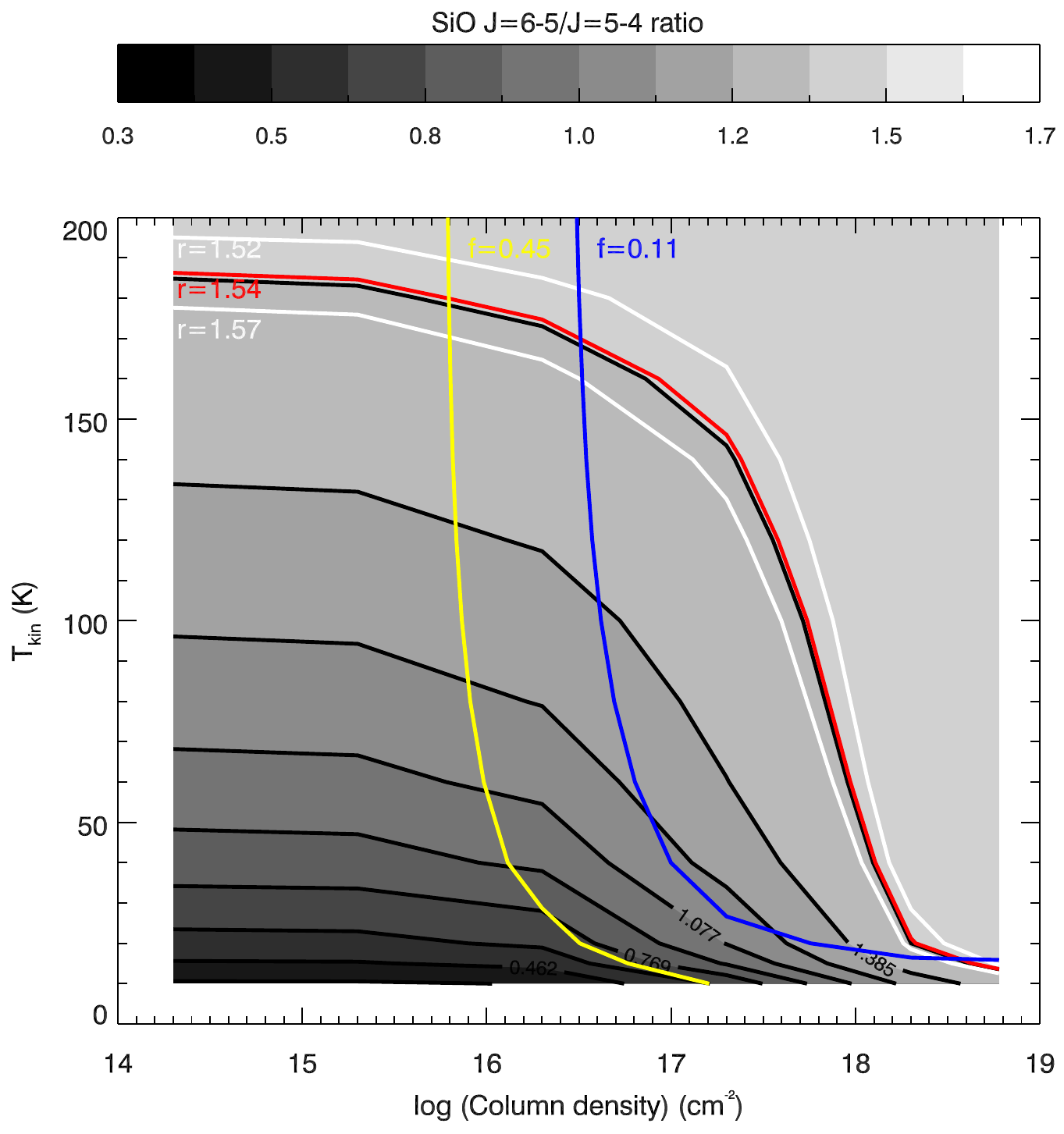}}
\caption{ The {\sc RADEX} models of the SiO line intensity ratio $r=I$(SiO $J$=6--5)/$I$(SiO $J$=5--4)
at $n_{coll}=10^6$\,cm$^{-3}$. 
The red line shows the parameter space that matches the measured  line ratio  ($r=1.54$),
and the white lines indicate the uncertainties of the measured line intensity ratio ($r=1.54^{+0.3}_{-0.2}$).
The right side of the blue and yellow lines shows the
parameter space where the thresholds set by the filling factor (likely limit of $<0.11$ and  probable limit of $<0.45$) are satisfied.
The solution for SiO modelling is the region where both the line ratio ($r=1.54^{+0.3}_{-0.2}$)
and filling factor (likely limit of $<0.11$ and  probable limit of $<0.45$) are satisfied.
\label{fig-sio-ratio}}
\end{figure}
\begin{figure}
\centering
\resizebox{\hsize}{!}{\includegraphics{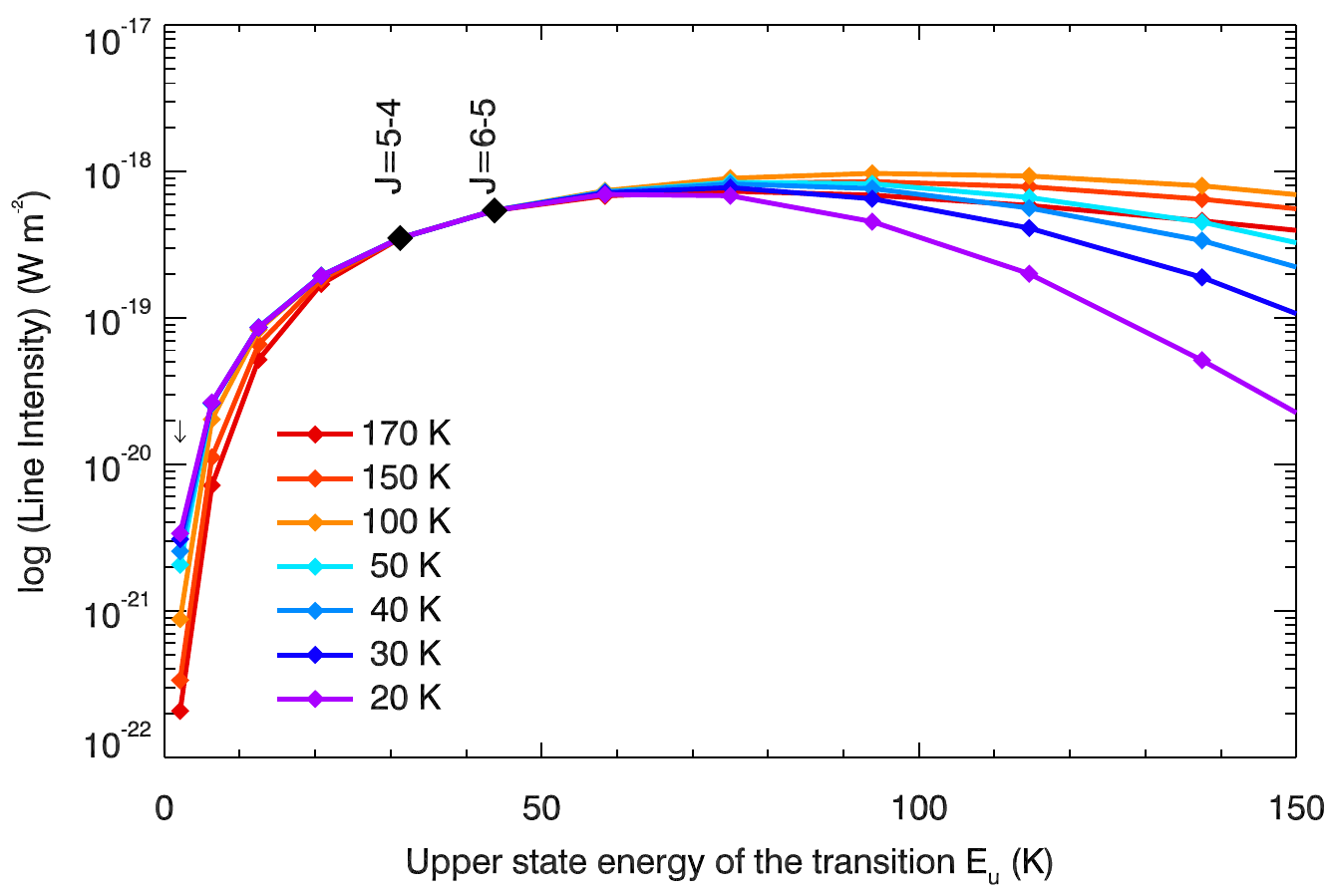}}
\caption{ The spectral line energy distribution (SLED) of SiO. The black diamonds
show the measured line intensities. The {\sc RADEX} models, plotted with  lines for seven different $T_{\rm {kin}}$ values,
show little difference  amongst models. The only difference is found 
at $J>$8--7, making difficult to constrain SiO kinetic temperature
from ALMA observations only.
\label{fig-sio-sled}}
\end{figure}

\begin{table}
 \caption{ SiO and CO model parameters \label{table-mass}}
\begin{center}
 \begin{tabular}{r ll l l} \hline
  $T_{{\rm kin}}$ & $n_{\rm coll}$ & $N$ & $f$ & $M$ \\ 
  K &  cm$^{-3}$ & cm$^{-2}$  & &  \Msun \\  \hline
SiO \\
170  & $10^{6}$ & $3.8\times10^{16}$  & 0.046  & $4\times10^{-5}$  \\
150  &          & $1.6\times10^{17}$  & 0.018 & $6\times10^{-5}$  \\
100  &          & $5.3\times10^{17}$  & 0.013  & $2\times10^{-4}$  \\
 50  &          & $8.8\times10^{17}$  & 0.019  & $4\times10^{-4}$  \\ 
 40  &          & $1.1\times10^{18}$  & 0.022  & $5\times10^{-4}$  \\
 30  &          & $1.3\times10^{18}$  & 0.028  & $9\times10^{-4}$  \\
 20  &          & $2.1\times10^{18}$  & 0.028  & $2\times10^{-3}$  \\  \hline
CO \\
  40 & $10^{6}$ & $1.9\times10^{19}$ & 0.06 & $2\times10^{-2}$ \\
  30 &          & $1.6\times10^{20}$ & 0.03 & $7\times10^{-2}$ \\
  20 &          & $1.3\times10^{21}$ & 0.05 & 1.0 \\
  50 & $10^{5}$ & $6.4\times10^{19}$ & 0.03 & $3\times10^{-2}$ \\
  40 &          & $1.8\times10^{20}$ & 0.02& $6\times10^{-2}$ \\
  30 &          & $4.7\times10^{20}$ & 0.03 & $2\times10^{-1}$ \\  \hline
 \end{tabular}\\
$T_{{\rm kin}}$: the kinetic temperature,
$N$: column density,
$f$: filling factor. \\
 \end{center}
\end{table}

We analysed the measured line intensities by modelling them with the non-LTE radiative transfer code {\sc RADEX} 
\citep{vanderTak:2007be}.
The gas density of the line emitting region may  not  be high enough for LTE to apply,
and a non-LTE radiative transfer code is needed for modelling.
We describe our modelling of the SiO lines in detail, followed by a brief description of similar analyses for other molecules.

\subsection{SiO analysis}

The {\sc RADEX} code calculates the level populations of molecules using collisional cross-sections.
We used SiO Einstein coefficients and SiO--H$_2$ collisional cross-sections from the {\sc LAMDA} molecular and atomic data base \citep{Schoier:2005ja}, which was based on the work of  \citet{Dayou:2006ey}.
Although H$_2$ has been detected in the inner ejecta \citep{Fransson:2016jb}, the main collisional partner of SiO may be  O$_2$, SO or O rather than H$_2$.
Faure (private communication) estimates that having O$_2$ as a collision partner could increase the rate coefficients by a factor of 1--10, compared with the H$_2$ collisional partner adopted here.
That may contribute to the uncertainties of the analysis.

To account for optical depth effects in emission lines we use the option LVG in the {\sc RADEX} code, which is equivalent to the Sobolev formalism for a freely expanding gas, as summarised by equations (7) and (8) of \citet{McCray:1993p29839}.

One of the input parameters is the gas density of the collisional partner.
We calculated line intensities by varying the gas densities of the collisional partner ($n_{\rm coll}$) from $10^4$, $10^5$ to $10^6$\,cm$^{-3}$.
We found that  a $10^5$\,cm$^{-3}$ model cannot achieve the measured SiO $J$=6--5 to 5--4 line ratio,
excluding the density being this value.
With $10^4$\,cm$^{-3}$ we found a solution that fitted both SiO measured line intensities, however, the required optical depth ($\tau$)
becomes more than 1000. 
For such a high optical depth in the SiO main isotope lines, the $^{29}$SiO isotopologue lines
would also become optically thick, so that 
 {\sc radex} predicts almost equally strong $^{28}$SiO and $^{29}$SiO lines
even with a very low isotope abundance ratio (see Sect.\ref{section-isotope}).
Because improbably strong $^{29}$SiO isotopologue lines were predicted, a density $10^4$\,cm$^{-3}$ can therefore be excluded.
Thus, we adopted the gas density of the collisional partner of $10^6$\,cm$^{-3}$ for the analysis.

Our adopted gas density is similar to  the density used by \citet{Jerkstrand:2011fz}.
They modelled SN\,1987A spectra taken eight years after the explosion, 
by solving for the thermal balance and radiative transfer.
The density of the O/Si/S zone was $4.2\times10^6$\,cm$^{-3}$.
Assuming a constant expansion of the clumps over time, that density would have been decreased to
$2\times10^5$\,cm$^{-3}$ twenty four years after the explosion.
Our adopted density is therefore more or less consistent to the value used by \citet{Jerkstrand:2011fz} with considering a constant expansion over time.

The other two input parameters are 
 the SiO column density ($N_{{\rm SiO}}$) and kinetic temperature ($T_{{\rm kin}}$).
Appropriate ranges of $T_{\rm {kin}}$
and $N_{\rm SiO}$ are constrained by the SiO $J$=6--5 to 5--4 ratio of integrated line intensities, 
$r=I$(SiO $J$=6--5)/$I$(SiO $J$=5--4).
Fig.\,\ref{fig-sio-ratio} demonstrates the set of ($T_{\rm {kin}}$, $N_{\rm SiO}$) values that match the ALMA measured
 line ratio ($r=1.54^{+0.3}_{-0.2}$) shown by a red line, with its uncertainties indicated by two white lines.

While the {\sc RADEX} models  line intensities refer to  whole sky emission ($4\pi$),
 the net line flux  from SN\,1987A is limited by its emitting surface area ($f\Omega$).
We define the filling factor ($f$) against the integrated area used for the spectrum extraction from the ALMA cube data
($\Omega=2.6\times10^{-11}$\,sterad).
The maximum possible size of the emitting area is limited by the estimated ejecta size.
There are several ways to constrain the ejecta size.
One  comes from the assumption that the ejecta gas is expanding with a constant velocity over 27\,years.
Although our measured FWHM for the SiO lines is $\sim$2000\,km\,s$^{-1}$, slightly wider FWHMs
were detected  at early times  \citep[2400\,km\,s$^{-1}$; ][]{McCray:1993p29839}, and we adopt the latter value for the possible ejecta size.
Choosing an equation converting the FWHM to the expansion velocity for a uniform sphere \citep{McCray:1993p29839},
 our measured FWHM corresponds to an expansion velocity of 1700\,km\,s$^{-1}$.
This gives an approximate emitting area of $2.8\times10^{-12}$\,sterad (0.19\,arcsec in radius)
after 27 years, setting the beam filling factor to be 0.11 or smaller. 
We adopt an LMC distance of 50$\pm$0.19$\pm$1.11\,kpc \citep{Pietrzynski:2013ck}, which is consistent with the distance measured towards
SN\,1987A (51.2$\pm$3.1\,kpc) by \citet{Panagia:1991io}.
The higher velocity component  was recorded to be 3500\,km\,s$^{-1}$  \citep{McCray:1993p29839}, though it was recorded from hydrogen
recombination lines.
The  longest axis of the H$\alpha$ image in 2012 is about 0.8\,arcsec  \citep{Larsson:2013gx, Fransson:2015gp}.
These give a maximum possible filling factor of 0.45.
In summary, we set $f<0.11$ as the more likely range, and $<0.45$ as a potential maximum value.

By comparing the measured line intensity of SiO $J$=5--4 with the modelled value,
we obtain the filling factor $f$. 
In Fig.\ref{fig-sio-ratio}, the contours for $f=0.11$ and $f=0.45$ are indicated by blue and yellow lines,
showing that only the  right top side of the pair of ($T_{\rm {kin}}$,
 $N_{\rm SiO}$) can fulfil the $f<0.11$  or $<0.45$ conditions.
The combination of the line ratio and the intensity gives constraints on $T_{\rm {kin}}$, $N_{\rm SiO}$ and $f$,
and the appropriate range for these parameters are summarised in Table\,\ref{table-mass}.

Figure\,\ref{fig-sio-sled} shows the energy diagram of SiO.
There are multiple solutions which fit the measured SiO line intensities,
and there is little difference in the SiO energy distributions between the $T_{\rm {kin}}$=20--170\,K models, 
except for $J$=1--0, 2--1 and $J>$8--7.
With existing ALMA data only, it is difficult to constrain the parameter space even further. 

The obtained column density is an `averaged' value within the beam, and is converted to the total molecular mass by $M_{\rm SiO}=f\Omega N_{\rm SiO}  d^2 m$, where  $d$ is the distance to SN\,1987A (50\,kpc),  and $m$ is the molecular mass.
The estimated SiO masses are listed in Table\,\ref{table-mass}.
The appropriate  range of the kinetic temperature  is 20--170\,K and the resulting SiO mass range is $4\times10^{-5}$--$2\times10^{-3}$\,\Msun.
Note that the inferred SiO mass is much smaller than that of CO,  despite their comparable line intensities.
This is so because SiO has a dipole moment $\sim$30 times larger than that of CO, hence an Einstein A coefficient $\sim10^3$ times larger.

\subsection{CO analysis}

\begin{figure}
\centering
\resizebox{0.98\hsize}{!}{\includegraphics{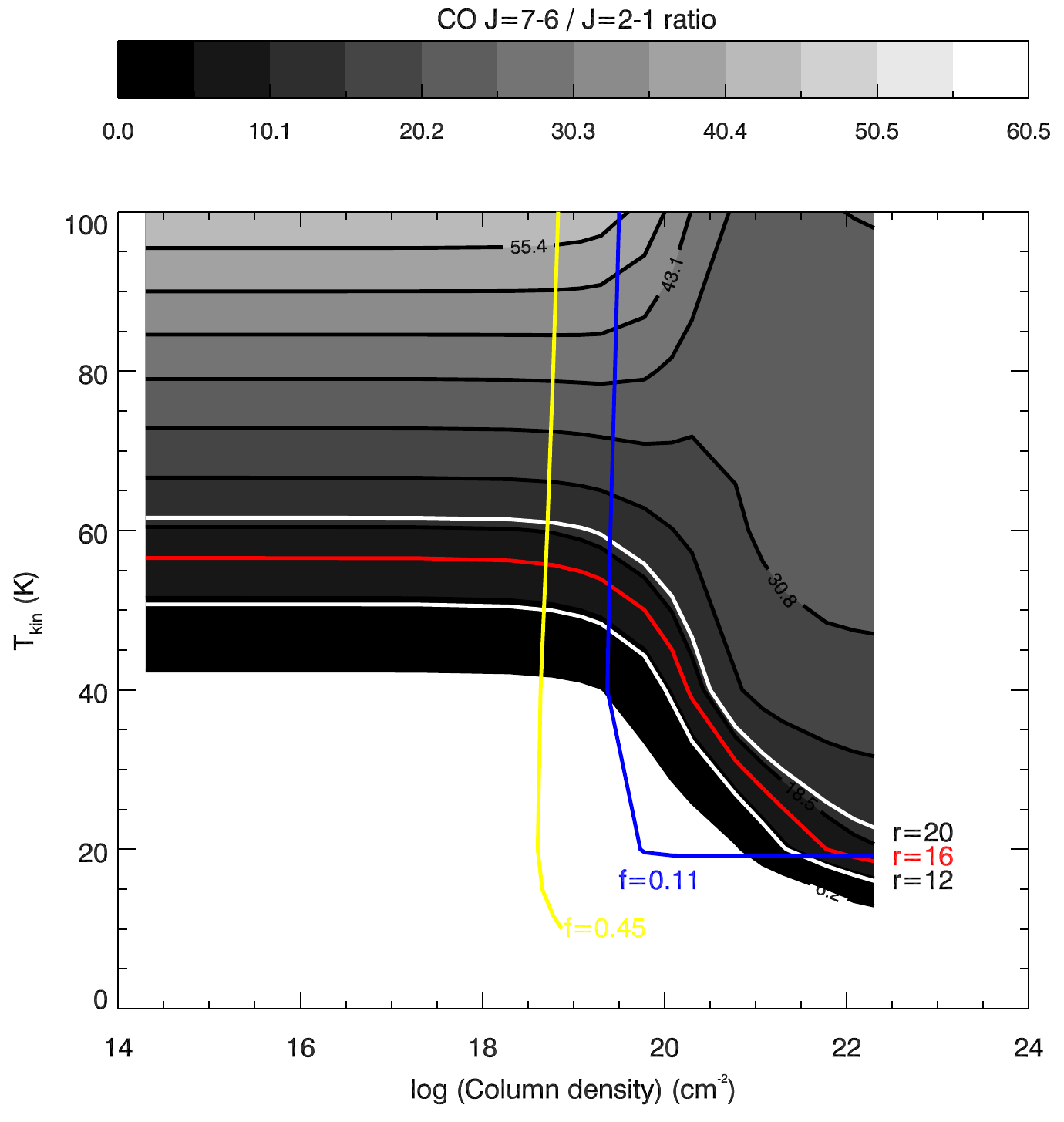}}
\resizebox{0.98\hsize}{!}{\includegraphics{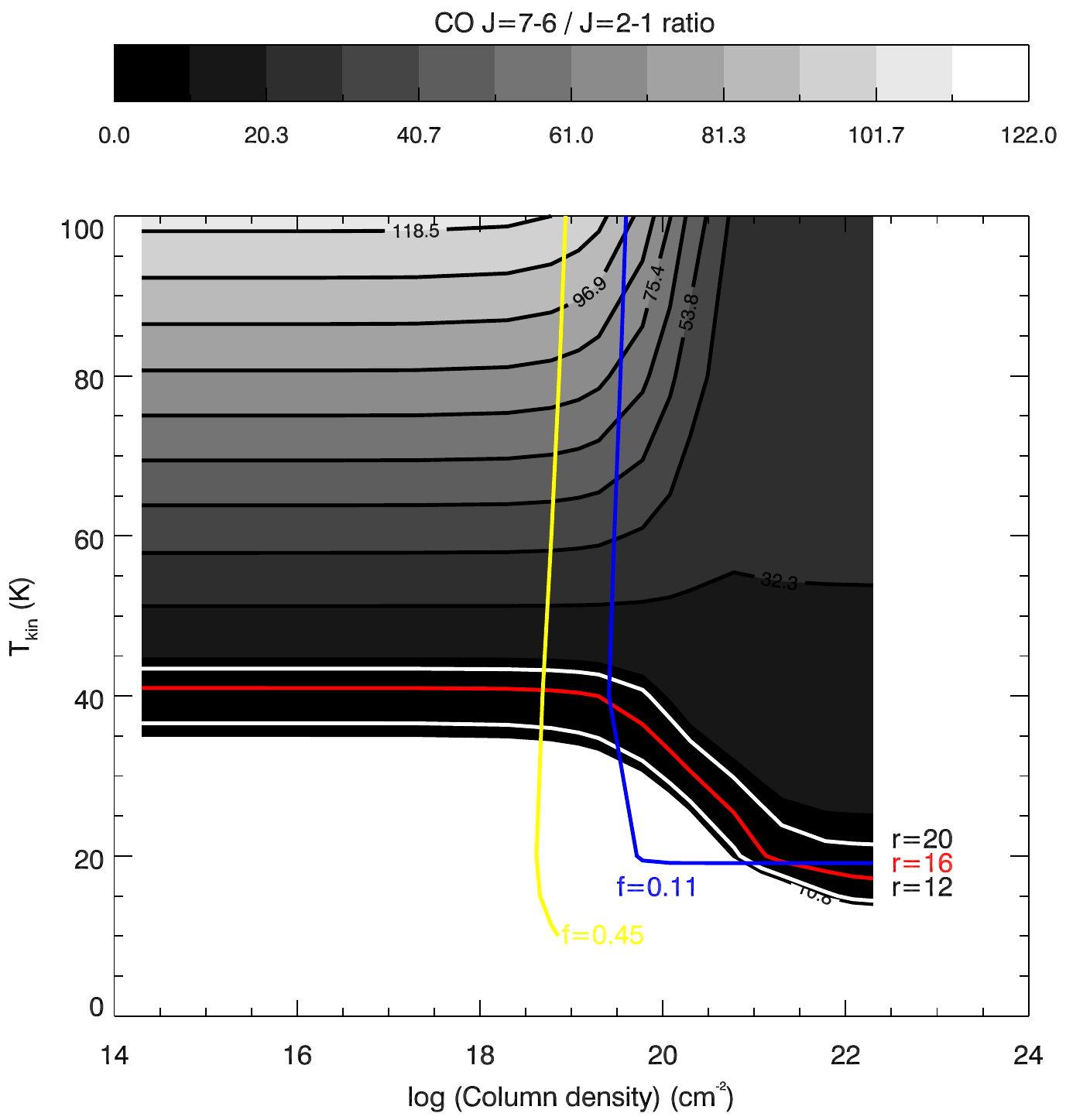}}
\caption{ {\sc RADEX} models for the CO line ratios ($r$=$I$(CO $J$=7--6)/ $I$ (CO $J=$2--1))  at $n_{\rm coll}$=$1\times10^5$\,cm$^{-3}$ (top)
and $n_{\rm coll}$=$1\times10^6$\,cm$^{-3}$ (bottom). 
The white lines show the range that fits the measured line intensity ratio ($r$=16), with its range of uncertainties
shown as grey lines ($r$=20 and 12).
The blue contour shows the filling factor threshold (f$<$0.11), and the potentially feasible range  ($<$0.45; yellow).
The cross section from these contours ($r=16\pm4$ and $f<0.11$) is the suitable parameter range for the CO fitting,
and the other cross section ($r=16\pm4$ and $f<0.45$) is the potential parameter range.
\label{fig-co-ratio}}
\end{figure}
\begin{figure}
\centering
\resizebox{\hsize}{!}{\includegraphics{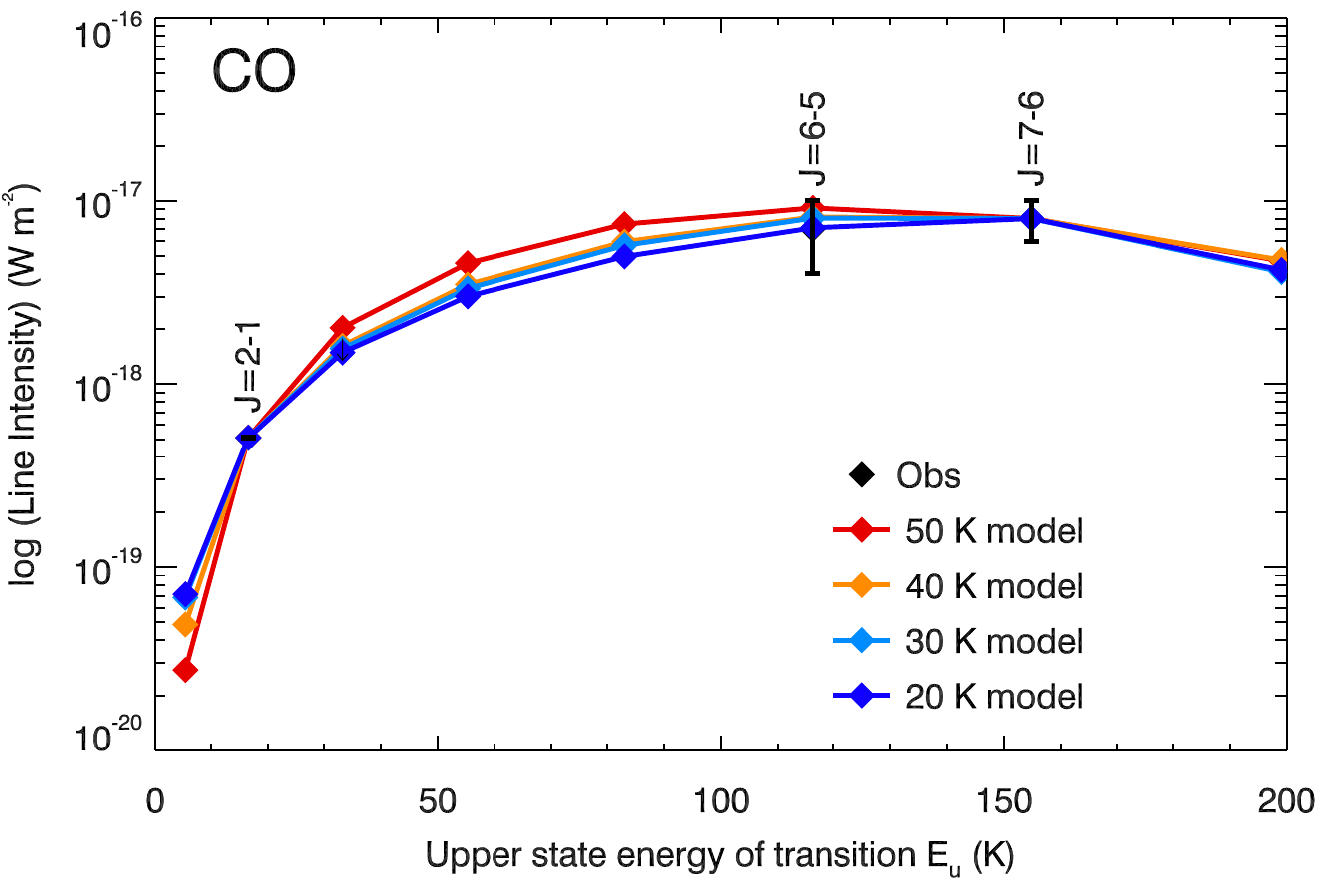}}
\caption{ The CO SLED from the ALMA and {\sc Herschel} measurements,
compared with $n_{\rm coll}$=$10^5$\,cm$^{-3}$ {\sc RADEX} model.
The models for $n_{\rm coll}$=$10^6$\,cm$^{-3}$ are almost the same as 
$n_{\rm coll}$=$1\times10^5$\,cm$^{-2}$  for a given temperature, so not plotted here.
\label{fig-co-sled}}
\end{figure}

The CO models have been revised after \citet{Kamenetzky:2013fv}, as the ALMA CO $J=$2--1 line flux has been re-calibrated.
Essentially, the CO analysis also follows the method used for SiO.
We used our latest CO $J=$2--1 measurements, together with published {\sc Herschel Space Observatory} $J$=7--6 and 6--5 measurements 
\citep{Kamenetzky:2013fv, Matsuura:2015kn}.
{\sc Herschel} CO fluxes were measured two years before the ALMA CO line fluxes.
While some flux evolution might have occurred,
however, considering only small reduction (3\,\%) of $^{44}$Ti heating over two years
and small change in volume ($\sim$10\%),
we believe that the evolution of fluxes should be within the measured uncertainties.
We use  two constraints to find the best parameter space, the line ratio of $J$=7--6 
($8\times10^{-18}\pm2\times10^{-18}$\,W\,m$^{-2}$) to $J$=2--1 ($51.0\pm0.4\pm0.3\times10^{-20}$\,W\,m$^{-2}$),
and the threshold of the filling factor ($f<0.11$), with a possible range up to $f<0.45$.
Here, the filling factor is defined differently from \citet{Kamenetzky:2013fv}'s definition;
they defined a unity  filling factor 
as the size of the expanding gas over 24\,years with a constant velocity of 2000\,km\,s$^{-1}$,
whereas this work adopts the ALMA spectral-extracted aperture size, which is a 1.29''$\times$1.10'' ellipse.
We used the collisional cross sections and Einstein coefficients for CO--H$_2$ from \citet{Yang:2010bb}.

Figure\,\ref{fig-co-ratio} shows the contours of the model line ratios.
In the case of CO, both $n_{\rm coll}$=$1\times10^5$\,cm$^{-3}$ and $1\times10^6$\,cm$^{-3}$
can provide solutions to fit the CO lines. We present contours for 
$n_{\rm coll}$=$1\times10^5$\,cm$^{-3}$ and $1\times10^6$\,cm$^{-3}$, separately.
The ALMA and {\sc Herschel} line ratio of CO $J$=7--6 to 2--1 is plotted using the red line ($r=16\pm4$),
with its uncertainty given by the white lines.
The region within the two white lines indicates the suitable parameter space to fit the measured CO line ratio.
The  filling factor gives additional constraints.
In Fig.\,\ref{fig-co-ratio}, the parameter space which satisfies
the most likely range of $f<0.11$ and the possible range up to $f<0.45$
are plotted by blue and yellow lines.

The combination of the {\sc RADEX} parameter space that satisfies both the line ratio and the filling factor gives the solutions of the CO modelling.
The suitable range at $n_{\rm coll}$=$1\times10^5$\,cm$^{-3}$ is
$T_{\rm kin}$=30--50\,K with a CO mass of 0.2--0.03\,\Msun.
At $n_{\rm coll}$=$1\times10^6$\,cm$^{-3}$, the range is $T_{\rm kin}$=20--40\,K and a mass is 1.0--0.02\,\Msun.
The fitted parameters and resultant CO masses are summarised in Table\,\ref{table-mass}.

Fig.\,\ref{fig-co-sled} shows the energy diagram of CO lines.
The models that fit the CO $J$=7--6 and 2--1 line intensities can also fit CO $J$=6--5
line intensities as well.

\subsection{Estimating the masses of HCO$^+$, SO, CS and SiS}

We also estimated the masses and their upper limits for HCO$^+$, SO, CS and SiS.
We adopted a gas density of $10^6$\,cm$^{-3}$, consistent with the model solutions for both SiO and CO.
The mass  depends only slightly on the assumed $n_{\rm coll}$.
With $n_{\rm coll} =10^5$\,cm$^{-3}$,
the inferred mass increases by a factor of $\sim$3, compared with the mass for $10^6$\,cm$^{-3}$.
We also fixed the kinetic temperature of the gas to be 40\,K
because both the $^{12}$CO and $^{28}$SiO modelling gave solutions at this temperature.
These molecular lines are optically thin,
and the column density and the filling factor are inversely-correlated.
Strictly speaking, this inverse-correlation is not applicable in some non-LTE conditions, but within the range of temperature, column densities and 
collision partner densities,
this condition holds.
Hence, the filling factor is fixed at 1, leaving only  the column density
as the relevant parameter for the line intensities.
We assume that the 267\,GHz line is due to 
HCO$^+$ only, with no SO$_2$ contamination, giving an upper limit for  the HCO$^+$ mass.
The resulting masses and upper limits are summarised in Table\,\ref{table-mass2}.

\subsection{Isotopologue  ratios}

From the measured line intensities of $^{28}$SiO and $^{12}$CO
and the upper limits for $^{13}$CO, $^{29}$SiO and $^{30}$SiO, we can estimate the 
$^{12}$CO/$^{13}$CO, $^{28}$SiO/$^{29}$SiO  and  $^{28}$SiO/$^{30}$SiO isotopologue abundance ratios,
which can be assumed to be equal to the isotope ratios.
The $^{29}$SiO lines are blended with SO lines, so that only upper limits for $^{29}$SiO line intensities are available.
We assume that the excitation temperatures of $^{28}$SiO, $^{29}$SiO and $^{30}$SiO  are identical.
The isotopologue ratios
can be directly obtained from the ratio of the line intensities without {\sc RADEX} modelling,
assuming that all the lines are optically thin, and that all transitions are in LTE.
This method gives $^{28}$SiO/$^{29}$SiO$>$13, $^{28}$SiO/$^{30}$SiO$>14$
and $^{12}$CO/$^{13}$CO$>21$.

The {\sc RADEX} modelling can further constrain the isotopologue ratios,
because level populations can be in non-LTE, and 
because $^{28}$SiO can be getting optically thick at lower kinetic temperatures.
We assume that  $^{28}$SiO, $^{29}$SiO and $^{30}$SiO have the same kinetic temperature, giving a lower limit
to the $^{28}$SiO/$^{29}$SiO ratio.
The lower limit for $^{28}$SiO/$^{29}$SiO is  $>34$ at 170\,K,
but increases to $^{28}$SiO/$^{29}$SiO$>128$ at 50\,K,
and $^{28}$SiO/$^{29}$SiO$>163$  at 20\,K.
The ratios are summarised in Table\,\ref{table-isotope-mass}.
The lower limit for $^{28}$SiO/$^{30}$SiO is  $>$43 at 170\,K,
and increases to $>$188 at 50\,K,
and $>$401 at 20\,K.
The 170\,K lower limits for the isotope ratio from RADEX are lower values than the estimates from the line ratios only,
assuming the LTE and an optically thin case. This is because of the non-LTE effects.

As long as the isotopologues of the same species have emitting regions with a similar gas density
and a similar kinetic temperature, 
the isotope ratios are much more robust than the molecular masses
which are more strongly dependent on the adopted  kinetic temperature.

\begin{table}
 \caption{ Estimated masses and upper limits for HCO$^+$, SO, SiS, and CS, assuming
 $T_{\rm {kin}}$=40\,K and $n_{\rm coll}$=$10^{6}$\,cm$^{-3}$  \label{table-mass2}}
\begin{center}
 \begin{tabular}{ll l l l} \hline
Molecules  &  Mass \\ 
           &  \Msun \\  \hline
HCO$^+$    & $\leq5\times10^{-6}$ \\
SO         & $\sim4\times10^{-5}$ \\ \hline
SiS        & $<6\times10^{-5}$ \\
CS         & $<7\times10^{-6}$ \\
\hline
 \end{tabular}\\
 \end{center}
\end{table}
\begin{table}
 \caption{ Isotopologue ratio limits. \label{table-isotope-mass}}
\begin{center}
 \begin{tabular}{ll l l l} \hline
Isotopologues  & \multicolumn{2}{l}{Method} & Abd ratio \\ \hline
$^{12}$CO/$^{13}$CO   & \multicolumn{2}{l}{line ratio} & $>21$  \\
                      & {\sc RADEX} &  40\,K, $n=1\times10^{6}$\,cm$^{-3}$  & $>40$ \\
\hline
$^{28}$SiO/$^{29}$SiO & \multicolumn{2}{l}{line ratio} &  $>13\pm2$ (J=5--4) \\ 
                      &  \multicolumn{2}{l}{} &   $>30\pm10$ (J=6--5)   \\
                      & {\sc RADEX} & 170\,K, $n=1\times10^{6}$\,cm$^{-3}$ & $>34$ \\
                      &       & 150\,K & $>52$ \\
                      &       & 100\,K & $>100$ \\
                      &       &  50\,K & $>128$ \\
                      &       &  40\,K & $>137$ \\
                      &       &  30\,K & $>145$ \\
                      &       &  20\,K & $>163$ \\
\hline
$^{28}$SiO/$^{30}$SiO & \multicolumn{2}{l}{line ratio} & $>14$ \\ 
                      & {\sc RADEX} & 170\,K, $n=1\times10^{6}$\,cm$^{-3}$& $>43$ \\
                      &       & 150\,K & $>64$ \\
                      &       & 100\,K & $>112$ \\
                      &       &  50\,K & $>188$ \\
                      &       &  40\,K & $>227$ \\
                      &       &  30\,K & $>279$ \\
                      &       &  20\,K & $>401$ \\
\hline
 \end{tabular}\\
 Abd ratio: abundance ratio
 \end{center}
\end{table}

\subsection{Modelling of the CO and SiO line profiles}

\begin{figure}
\centering
\resizebox{0.7\hsize}{!}{\includegraphics{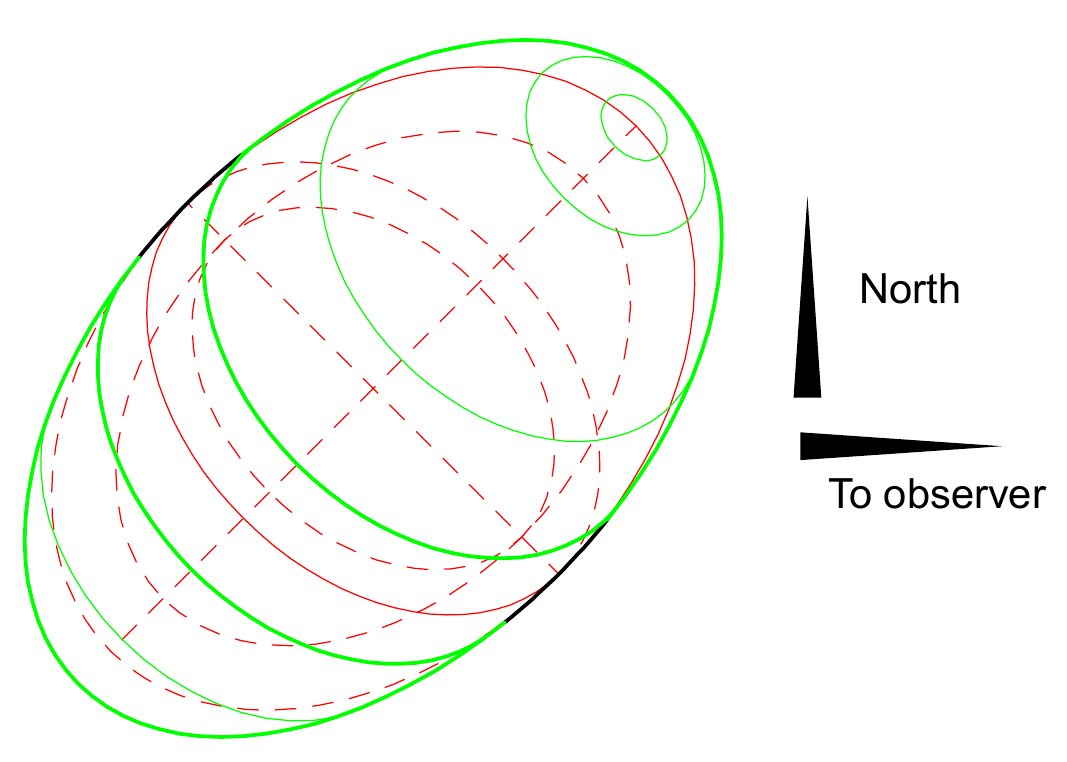}}
\caption{Schematic view of the SiO distribution used for velocity modelling.
Red lines show the poles and equator of the outer and  inner ellipsoids,
and the green lines indicate the SiO-rich region
\label{fig-lineprofile-in}}
\end{figure}

\begin{figure*}
\centering
\resizebox{0.7\hsize}{!}{\includegraphics{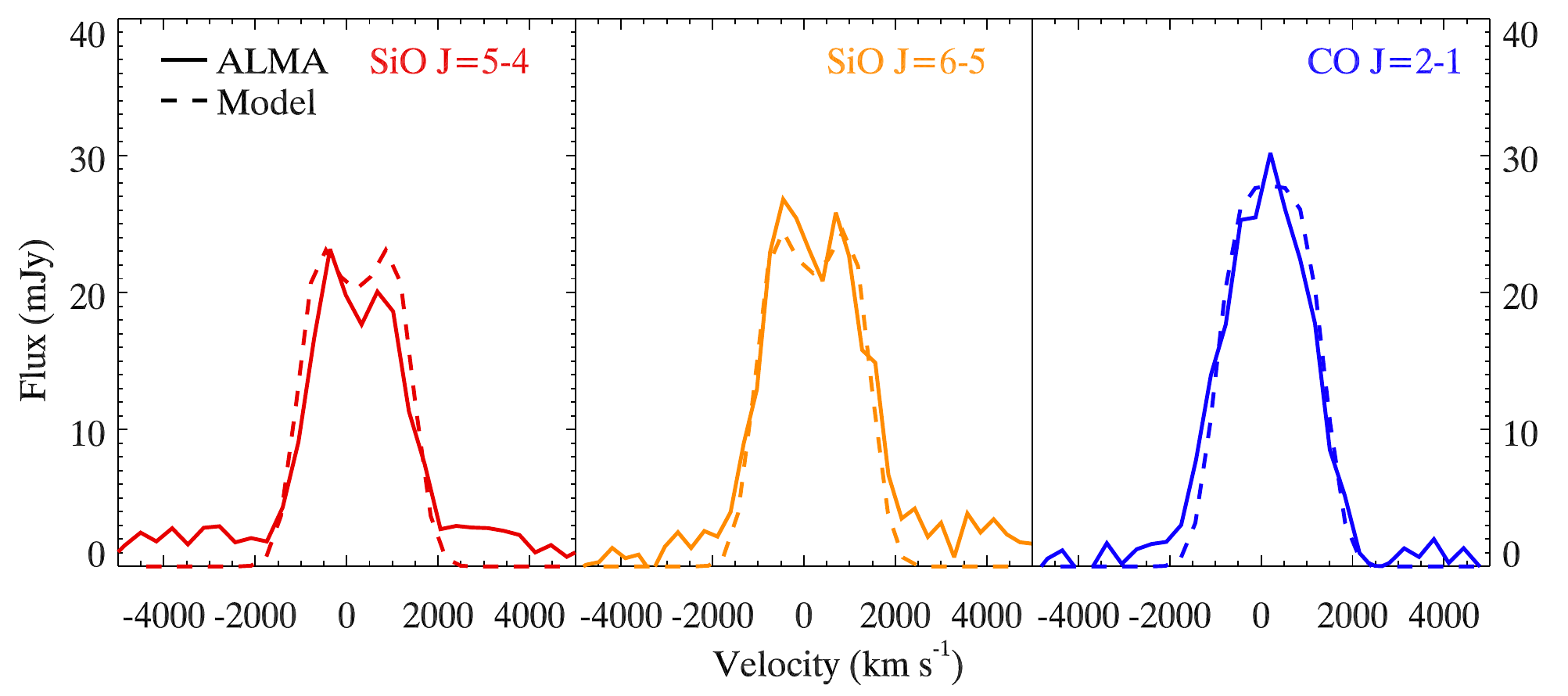}}
\caption{Model line profiles for CO  and SiO, compared with the observed ones.
The difference between the CO and SiO line profiles can be explained by morphological differences
(see text).
\label{fig-lineprofile}}
\end{figure*}

We present a very simple modelling of the line profile shapes
expected for the observed CO and SiO lines. The purpose of this section
is to show that the line profiles are compatible with reasonable ideas
about the line-emitting region and to deduce some general characteristics
of the molecular emission. The excitation requirements to
obtain the observed line intensities were discussed in Sect.\,\ref{analysis},
therefore here we focus on modelling the profile shapes, to be compared
with the observed ones.

Consistent with the observed shape of the molecular emitting region, 
we assume that the line-emitting region is a hollow,
prolate ellipsoid. In our toy-model, we take the ratio of the ellipsoid
axes to be  about 2; the inner radius of the ellipsoid is assumed to
be about 80\% of the outer one. The inclination of the long axis
with respect to the plane of the sky is taken to be of about 45
degrees. Note that the absolute sizes of the axes are not relevant at
this stage, when we discuss only normalised profiles, but we are
assuming that the emitting region is similar to the observed central
ellipsoid (Fig. \ref{fig-ch}). In order to explain the difference between the CO
and SiO profiles, we assume that the CO  emission comes from the whole ellipsoid, while SiO is abundant in the outer region
of the ellipsoid, with a significantly lower abundance at the inner region at 20\% of the long axis as illustrated in Fig.\,\ref{fig-lineprofile-in}.
As discussed in
Sect.\,\ref{analysis}, it is probable that the emitting region is in fact composed of
a large number of small clumps or filaments, occupying a small fraction
of the total volume. The above elliptical shape therefore applies just
to the overall large-scale distribution of the gas emitting in
the molecular lines, although the emitting gas, due to chemical and/or
excitation phenomena,  would be in fact concentrated in a large number of
clumps within the large-scale structure. As we will see, our
profile calculations are independent of the substructure of the mass
distribution forming many clumps, provided that they are homogeneously
distributed on large scales.

Since the line emitting gas is expected to have  been ejected in a
single very short event from a compact object, with minor
acceleration/deceleration after the explosion, ballistic radial
expansion is expected. We therefore assume radial expansion with a
linear velocity gradient. The fastest velocity (at the tip of the
ellipsoid) is taken to be about 1500\,km\,s$^{-1}$, in order to reproduce the
observed profile width.


For the assumed velocity field, the emission at each local standard of rest (LSR) velocity comes
from the flat region defined by cuts of the ellipsoid parallel to the
plane of the sky, in a kind of tomography of the emitting region. For
velocity channels with the same width, the distance between the planes
is constant. This property is not affected by the fact that the
line-emitting gas is placed in a large number of clumps within
the ellipsoid. We assume the excitation and abundance
distributions to be constant in the ellipsoid (at large scale),
except for the lack of SiO in equatorial regions; 
these simplified constraints are sufficient to reasonably
reproduce the observed profiles. 
The predicted emission from a shell showing a ballistic velocity field, in particular its relation with the size of the emitting regions at each velocity, was already discussed for some evolved nebulae with intense molecular lines e.g. 
by \cite{Bujarrabal:1997wo, Bujarrabal:1998tu, Bujarrabal:2007bd}.
Simple excitation conditions, geometry, and kinematics were also assumed in those papers, with predictions in good agreement with observations.  Our modelling is similar to that presented in those papers, except for the clumpy nature of the shell, which is an intrinsic, distinct property of  SN\.1987A. 
In the optically thick case (and for a
constant excitation temperature), the angle-integrated emission is
obviously proportional to the surface offered by the emitting region;
in the optically thin case, the intensity will be proportional to that surface
multiplied by the depth of the region emitting at each velocity, which
is also constant. In both cases, the brightness distribution for a
given velocity must also be constant across the angular extent of the
corresponding section. 
The final spectra were smoothed to a resolution of 375\,km\,s$^{-1}$, to account for the presence of a local velocity dispersion, 
which can be due to micro-turbulence or a dispersion in the velocities of different clumps, and to match the resolution of the smoothed ALMA spectra, 
which is of this order.
The modelling was motivated by the emission line profiles and consistent with recently obtained
high resolution imaging (Abellan et al. in preparation).

With these simple assumptions, we have derived 
normalised line profiles, which are comparable to the
observed ones within the uncertainties  (Fig.\,\ref{fig-lineprofile}). 
The molecular
emission might originate from a region having similar
geometrical and dynamical properties compatible with our assumptions.

\section{Discussion}

\subsection{Molecular chemistry}

It has been thought that supernovae are molecule-poor environments,
because hydrogen, the key element for chemical reaction, is expected to be rather deficient in SN ejecta,
and because ample He$^+$  destroys molecules formed within it
\citep[e.g.][]{Lepp:1990cz, Rawlings:1990th, Cherchneff:2011vq}.
For young ($<$10\,years old) SNe, CO and SiO had been the only molecules firmly detected
 \citep[e.g.][]{Kotak:2006p13473, Cherchneff:2011vq}, 
 with a recent report of detecting near-infrared H$_2$ lines from SN\,1987A since 2004 \citep{Fransson:2016jb}.
Now from our spectral line survey  with  ALMA, 
we have discovered SO and HCO$^+$ in the ejecta  of SN\,1987A. 
These detections add to the already  detected molecules at millimetre wavelengths, CO and SiO, and likely $^{29}$SiO,
\citep{Kamenetzky:2013fv}.
The ejecta of young SN remnants may represent at a much more molecule-rich environment than  had been thought before.
All the molecules found in SN\,1987A have formed from elements ejected and synthesised from this supernova, except for H$_2$,
and formed only after the explosion within the last $\sim30$\, years.
The inner ejecta have not yet interacted with circumstellar material from the progenitor star or ambient ISM material.

Supernova remnants that are over a thousand years old  have shown
slightly more diverse molecular species.
H$_2$O, OH$^+$, HCO$^+$ and SO molecules, in addition to CO, have been detected in SN remnants that are interacting with  molecular clouds 
\citep{vanDishoeck:1993ta, Reach:1998fw, Paron:2006iw}.
These molecules are thought to originate from the molecular clouds  rather than the SN remnants \citep{vanDishoeck:1993ta},
or from post-reverse shock regions \citep{Reach:1998fw, Wallstrom:2013he, Biscaro:2014kh}.
The only molecular detections which are not associated with shocks
are  H$_2$, ArH$^+$ and OH$+$ in  the Crab Nebula SN remnant \citep{Graham:1990kg, Loh:2012eu, Barlow:2013dr}.
 Due to  different  time scales and densities, as well as the presence of shocks, molecules in old SN remnants may
 have experienced different chemical pathways from young supernova remnants, such as SN\,1987A.

\subsubsection{History of molecular detections in SN\,1987A}

In the early epochs of SN\,1987A, all the CO and SiO detections were of vibrational bands at near- and mid-infrared (IR) wavelengths
\citep{Rank:1988hi, Spyromilio:1988hk, Aitken:1988wo, Catchpole:1988wz, Roche:1989jb}.
CO emission was found at day 192 at 2.3 and 4.6\,$\mu$m \citep{Rank:1988hi, Spyromilio:1988hk}.
A possible trace of SiO emission was reported at  day 257 \citep{Aitken:1988wo}, which 
was later confirmed by \citet{Roche:1989jb}. SiO continued to be present at day 517 \citep{Roche:1991vv}.
Both the SiO and CO emissions had disappeared from IR spectra by day 615 \citep{Wooden:1993p29432}.
Since that time, there had been  no SiO and CO detections in SN\,1987A for nearly 25 years.
Our  ALMA observations show CO and SiO rotational transitions  in millimetre spectra at day 10,053.
The emission has shifted from infrared to millimetre wavelengths on a time scale of two decades,
because the excitation temperatures of the molecules have decreased.



Our analysis finds a current CO mass of  0.02--1.0\,\Msun.
The CO mass estimated at early times was rather uncertain.
\citet{Spyromilio:1988hk}
reported $10^{-5}$--$10^{-4}$\,\Msun\, of CO at day 100,
while from the analysis of the same IR spectra 
\citet{Liu:1995jk} suggested 0.45\,\Msun\, at day 100, decreasing to 
(2--6)$\times10^{-3}$\,\Msun\, at days 200--600.
The difference was caused by the latter work accounting for the optical depth and non-LTE effects for CO,
while the former assumed optically thin emission with LTE assumed.
Nevertheless, our analysis shows that the current CO mass is larger than at day 200--600.
Such a large current CO mass may suggest that the destruction of CO via fast moving electron collisions has decreased in the last two decades.
This is reasonable, because the two key contributions to Compton electron production \citep{Thielemann:1990p29472} are
from  $^{56}$Ni (half-life of 6.1 days)  and $^{56}$Co \citep[half-life of 77.1\,days;][]{Nadyozhin:1994ip} whose decay rates reduced with time.
The reduction of the destruction rate of CO \citep{Deneault:2006bh} opened up the possibility to sustain a larger mass of CO now.

Alternatively,  \citet{Sarangi:2013bj} found that CO is primarily formed in the outmost zone of the O-rich core 
by neutral-neutral reactions involving O$_2$ and C atoms and by radiative association reactions. 
The chemical model predicted that these processes built up an increasing CO mass from $10^{-4}$ \,\Msun, at day 100
to $\sim$0.2\,\Msun, at day 2000.

Our {\sc RADEX} models show that
the gas density of $10^6$\,cm$^{-3}$ is  high enough for CO up to $J$=4--3 to be nearly in  LTE,
though at $10^5$\,cm$^{-3}$, CO level population of all transitions is slightly offset from  LTE.
So the gas density of SN\,1987A is at a borderline of  LTE and non-LTE.

The current molecular temperature is consistent or slightly higher than the dust temperature.
The dust temperature measured in 2012 was 20--30\,K, depending on the dust composition
\citep{Matsuura:2015kn}. The kinetic temperatures of the molecules
are between 20--50\,K for CO and  20--170\,K for SiO.
It is possible that the gas and dust are not to collisionally coupled.

\subsubsection{HCO$^+$ as a tracer of dense gas and possible requirement of mixing}

The surprising detection of HCO$^+$ in SN\,1987A can inform us about the density of the ejecta.
HCO$^+$ is commonly used as a dense gas tracer in molecular clouds with moderate ionisation
\citep[e.g.][]{2014oma..book.....W}.
This is because its critical density ($n_{\rm cr}$) is much higher than that for more commonly found molecules, such as CO.
The critical density of HCO$^+$ $n_{\rm cr}$(HCO$^+$ $J$=3--2) is  (2--1)$\times10^6$\,cm$^{-3}$ at 50--100\,K, while
that of CO $n_{\rm cr}$(CO $J$=2--1) is only (3--2)$\times10^3$\,cm$^{-3}$ at 50--100\,K.
Our detection of HCO$^+$, which is as strong as one eighth of the CO line intensity,
 shows that the gas density in the HCO$^+$ emitting region
in the ejecta is far higher than (3--2)$\times10^3$\,cm$^{-3}$, and probably close to $\sim10^6$\,cm$^{-3}$,
consistent with the density found from our SiO and CO analysis.

As chemical models by \citet{Sarangi:2013bj}  did not consider HCO$^+$ formation, 
while \citet{Rawlings:1990th}  did consider it but  predicted very small HCO$^+$ mass,
we revisit the chemical processes to form this molecule.
Forming HCO$^+$ requires H$_2$, which is abundant only in the hydrogen envelope.
HCO$^+$ is formed via the reaction of CO with H$_3^+$,
($\mathrm{H_3^+ + CO \rightarrow H_2 + HCO^+}$), and
the formation of H$_3^+$ requires H$_2$ and a moderate ionisation state
($\mathrm{H_2^+ + H_2 \rightarrow H_3^+ + H}$) \citep{2014oma..book.....W}.
These reactions can occur both in oxygen-rich and carbon-rich environments.
In total, forming HCO$^+$  requires H$_2$ in CO gas.
The reaction also requires H$_3^+$, which was suggested to have been detected in the early days \citep{Miller:1992kw}.
Alternatively, the reactions $\mathrm{C + H_2 \rightarrow CH + H}$
and $\mathrm{CH + O \rightarrow HCO^+ + e^-}$ can form HCO$^+$ after $>$1000\,days
 \citep{Rawlings:1989tl, Rawlings:1990th}.
Both chemical processes require H$_2$ to form HCO$^+$.

A substantial mass of CO is required to form HCO$^+$ in the case of the reaction via CO.
The HCO$^+$/CO abundance ratio is density dependent.
The highest abundance of HCO$^+$ is usually found in dense regions,  where
the HCO$^+$/CO abundance ratio can be as high as 10$^{-4}$ \citep{Papadopoulos:2007db},
while  typically the abundance is far lower than $10^{-7}$ in the ISM \citep{Viti:2002jc}.
In order to form  $\sim5\times10^{-6}$\,\Msun\, of HCO$^+$, the gas should contain about  $10^{-2}$\,\Msun\, of CO.
There is a small fraction of C and O in the hydrogen envelope, because  C and O are present there from when
they were incorporated into the star at the time of its formation. 
However, the masses of intrinsic C and O are  insufficient to account for the CO mass needed to generate  substantial HCO$^+$ mass (only about $10^{-4}$\,\Msun\, of CO at most).
In contrast, such a high abundance of CO is available  in the inner ejecta, while a large mass of H$_2$ must have formed in the hydrogen-envelope.

The enigma of CO in the ejecta interior and H$_2$ in the hydrogen envelope might be solved by the presence of some macroscopic mixing in the ejecta in the early phases after the SN explosion.
A classic picture of stellar nucleosynthesis is that the progenitor undergoes a sequence of nuclear reactions in the stellar interiors, building multiple zones with discrete elemental abundances.
Figure\,\ref{fig-mixing} (a)
shows the radial distributions of elements in an unmixed case, starting from the innermost region (interior mass=0\,\Msun)  outwards.
This figure is based on explosive nucleosynthesis models for SN\,1987A \citep{Sukhbold:2016bo}, but includes a more extensive nuclear reaction network based on  \citet{Woosley:1988p29481}  and  \citet{Woosley:1997p29921} models.
The model is for an 18\,\Msun\, star with an initial metallicity of one third of the solar metallicity.
In the unmixed case, the hydrogen envelope does not have any benefit of enhanced oxygen and carbon atoms being synthesised in the stellar interior.
So the presence of some mixing could help in forming HCO$^+$.
Indeed, recently, H$_2$ emission has been detected in the inner ejecta of SN\,1987A \citep{Fransson:2016jb}, so one of the key molecules to form HCO$^+$ is present in the inner ejecta.

\begin{figure}
\centering
\resizebox{\hsize}{!}{\includegraphics{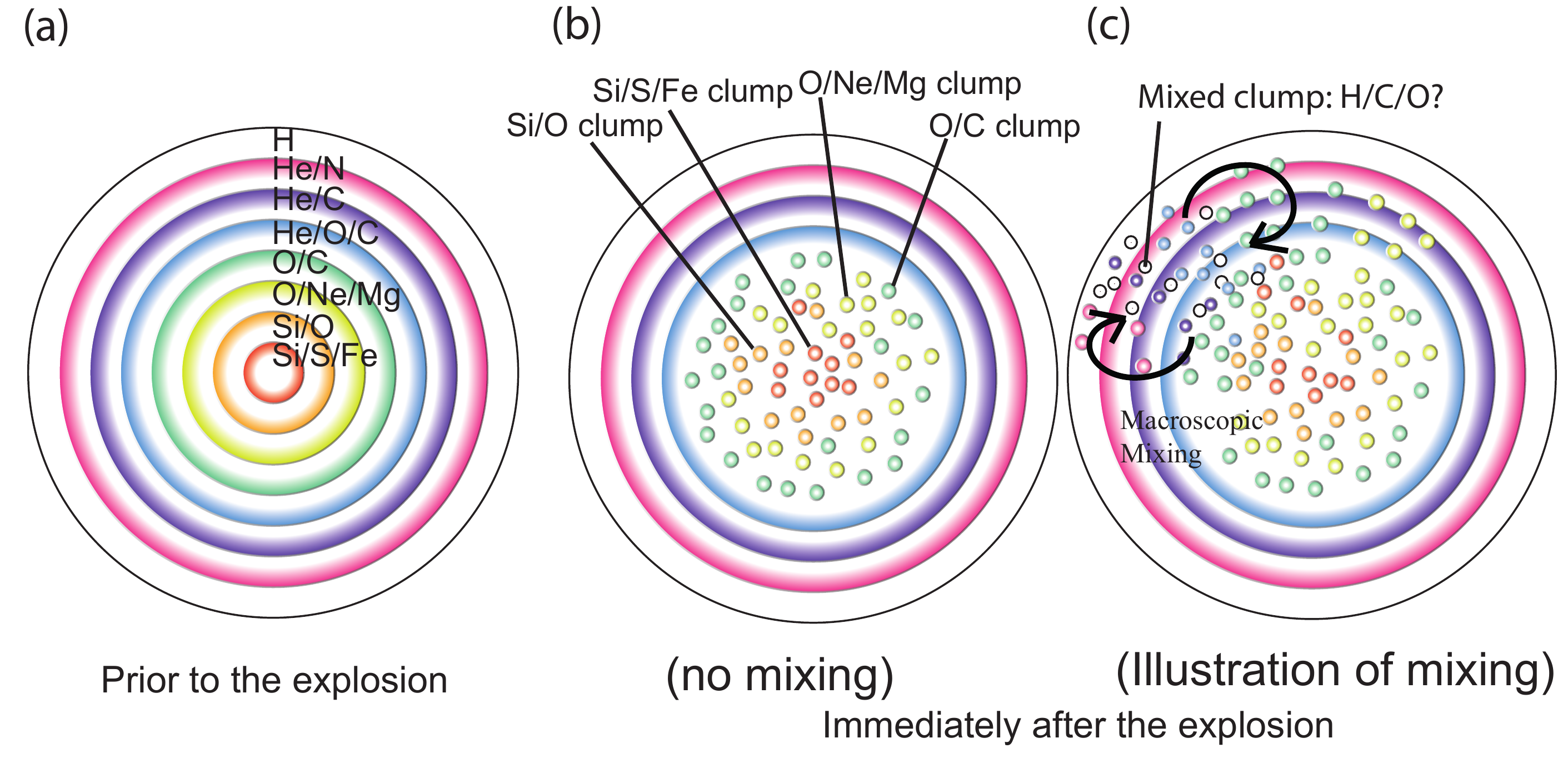}}
\caption{ Illustration of the effect of clumping and macroscopic mixings.
Panel (a) shows the `classic' picture of a stellar core prior to the explosion,
consisting of onion layers of different nuclear-burning shells.
Hydrodynamical simulations suggest that Rayleigh-Taylor instabilities break elemental zones into clumps (b).
Further, macroscopic mixing may make some clumps composed from multiple elemental zones (c).
 \label{fig-onion}}
\end{figure}

Recent hydrodynamical simulations have shown that
Rayleigh-Taylor instabilities  and Richtmyer-Meshkov instabilities mix elements
at the interfaces of He-envelope and C+O zone \citep{Kifonidis:2000bu, Hammer:2010di}.
The Rayleigh-Taylor instabilities  break elemental zones into clumps (Fig.\,\ref{fig-onion} b),
and Richtmyer-Meshkov  instabilities make some fraction of hydrogen sink into the inner ejecta zone  (macroscopic mixing; Fig.\,\ref{fig-onion} b).
This mixing can bring hydrogen atoms down, and elevate  oxygen and carbon atoms,
increasing the  carbon and oxygen abundance in hydrogen-rich regions and the hydrogen abundance in the C+O region.
In Fig.\,\ref{fig-mixing} (b), artificial mixing was introduced by disturbing the velocity gradient of 300\,km\,s$^{-1}$
that matches the light curve, $\gamma$-ray brightening and the spectrum of SN\,1987A 
\citep{Pinto:1988bg, Woosley:1988p29481, Arnett:1989p29666}.
This demonstrates an example of the mixing effect ---
 allowing some hydrogen to co-exist with carbon and oxygen rich ejecta.
Note that  Fig.\,\ref{fig-mixing} (b) is one dimensional model, which does not account for clumpiness,
which has been illustrated in Fig.\,\ref{fig-onion} b.

 If  mixing can enhance CO in H$_2$ gas, or H$_2$ gas can sink into the C+O zone,
this can potentially increase the HCO$^+$ abundance in the ejecta.
That brings enigma of  mixing --- what is the extent of the mixing and what types of mixing would be needed?
\citet{Fransson:1989ie} argued against microscopic mixing, because that would cause discrepancies between line ratios of ionised or neutral lines
observed in early phases ($<$200 days). However, they open the possibility of some extent of macroscopic mixing of clumps.
Indeed, hydrogen rich gas from the H envelope moving as slow as 400\,km\,s$^{-1}$ has been observed from H$\alpha$ and H$_2$ \citep{Kozma:1998p29913, Larsson:2016wg}, 
which could potentially sank into the inner core.
Instead of mixing elements completely,  mixing might occur locally at the interfaces of the He and He-envelopes and 
the He and C+O zone, due to Rayleigh-Taylor instabilities
\citep{Mueller:1991p29863}.
Depending on model inputs, Rayleigh-Taylor instabilities at  the C+O and He interface can merge into the H and He interface, developing a larger scale of mixing 
\citep{Herant:1991ed, Wongwathanarat:2015jv}.
The presence of a such large scale mixing at early times could have later resulted in a substantial mass of HCO$^+$.

The general assumption
is that the ejecta material becomes chemically less active, after the earlier phases (up to about 1000\,days) because both the temperature  and the density drop.
However, our estimated gas density for the ejecta is  of the order of $10^6$\,cm$^{-3}$ currently.
Even with a temperature as low as 25\,K, the Maxwellian velocity is about 10$^5$\,cm\,s$^{-1}$.
With this density, molecules can collide with their counter parts in less than fraction of a second,
so that the ejecta might be still chemically active.
Moreover, \citet{Rawlings:1990th} predicted that He$^+$ abundance should decrease at about day 1000\,days,
and consequently, molecular abundances would increase. Also at reduced density, neutral-neutral reactions,
which would require overcoming high energy barriers, would be inefficient, and instead ion-molecular reactions
may be taken place (Cherchneff, private communication).
These might have caused changes in the molecular compositions over the time.

\citet{Rawlings:1990th} published the only model that predicted the formation of HCO$^+$ in SNe.
The process required was $\mathrm{C + H_2 \rightarrow CH + H}$
and $\mathrm{CH + O \rightarrow HCO^+ + e^-}$.
The predicted fractional abundance at day$\sim$1000 was however only $10^{-18}$.
These models account for elemental abundances from the stellar yields, but not for nuclear burning zone structures.
Incorporating zone structures could potentially increase  the HCO$^+$ abundance.

\subsubsection{ SiO and SiS chemistry}

Chemical models predict that at day 1500 the majority of  SiO is depleted into silicate dust grains
\citep{Sarangi:2013bj, Sarangi:2015he}.
Our measured SiO mass is consistent with this hypothesis qualitatively, but not quantitatively.
Our estimated SiO mass is $4\times10^{-5}$--$2\times10^{-3}$\,\Msun\, which is a factor of 16--800 larger than
the theoretically predicted SiO mass  \citep[$2.5\times10^{-6}$\,\Msun; ][]{Sarangi:2013bj}.
Nevertheless, Si atoms in SiO is only $3\times10^{-5}$--$1\times10^{-3}$\,\Msun\,
which corresponds to less than 10\,\% of Si mass synthesised in SNe.
That leaves the possibility that a large fraction of Si could be in silicate dust. Alternatively, Si can be in atomic form, as found in [Si\small{I}]
\citep{Kjr:2010p29878}.

\begin{figure*}
\centering
\resizebox{\hsize}{!}{\includegraphics{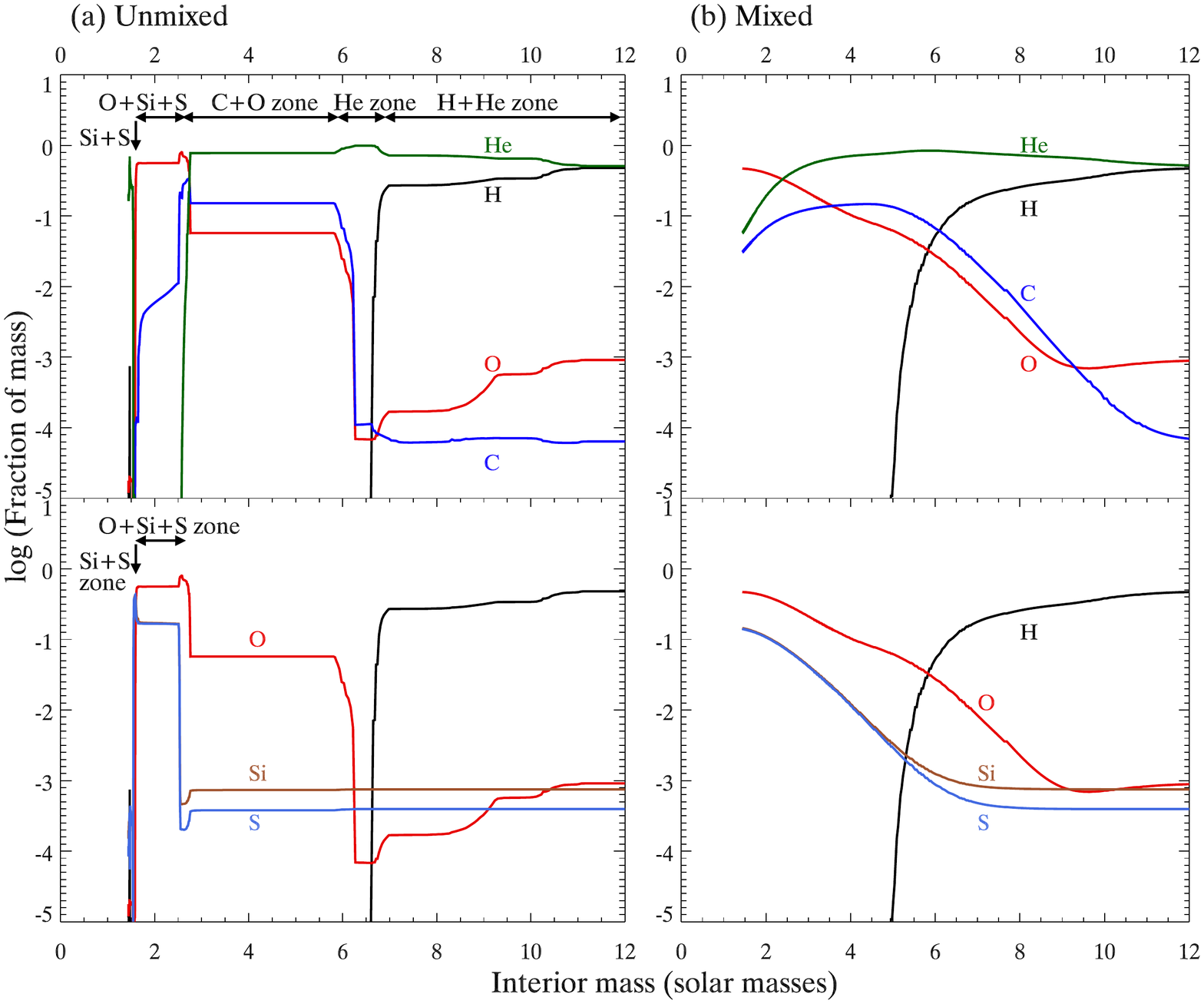}}
\caption{ The modelled fractional abundance of atoms after SN explosion.
The x-axis shows the interior stellar mass, with zero \Msun\, being the inner most.
The model is an 18\,\Msun\, star with one third of the solar metallicity  
based on the model for Sk$-$69$^{\circ}$\,9202 \citep{Sukhbold:2016bo}, but involving more extensive nuclear reaction network.
Only the inner 12 solar masses of the 17.09\,\Msun\, solar mass pre SN  are shown.
Panel (a) shows SN\,1987A model without mixing of nuclear burning zones.
Panel (b) shows an example of the effect of mixing,
with the model with artificial  mixing at $t=10^7$\,s.
For clarity, we split the plots in two (upper and lower panels).
The upper panel is for $^1$H, $^4$He, $^{16}$O and $^{12}$C, and the lower panel is for $^{28}$Si and $^{32}$S with H and O for guidance.
The unmixed model (a) shows distinct nuclear burning zones, Si+S, O+Si+S and C+O zones,
as well as  He and H+He envelopes.
(b) Having mixing triggered by the explosion can make elements from different nuclear burning zones mixed to some degree,
though the degree of mixing is largely in debate.
The most notable effect of mixing is  Si and S being mixed into O (right lower panel),
which was not the case in the unmixed model (left lower panel).
\label{fig-mixing}}
\end{figure*}

Chemical models \citep{Sarangi:2013bj} predict that the majority of Si is actually in the form of SiS rather than SiO.
This is because  the innermost zone (Si+S+Fe zone) of the stellar core is oxygen-deficient, 
and Si in this zone forms SiS. The predicted SiS mass is $4.4\times10^{-2}$\,\Msun.
However, we find only small mass of SiS ($<6\times10^{-5}$\,\Msun).
This discrepancy may be partly explained by input nucleosynthesis models.
The chemical models of \citet{Sarangi:2013bj} used elemental abundance
predicted for a 19\,\Msun\, SN progenitor with solar abundances from
\citet{Rauscher:2002p29812}, which is not optimised for SN\,1987A  nor LMC abundances.
The \citeauthor{Rauscher:2002p29812} nucleosynthesis model predicts a large Si+S+Fe zone mass of 0.11\,\Msun, 
where $4.4\times10^{-2}$\,\Msun\, of SiS can form.
In contrast, Fig.\,\ref{fig-mixing} (a), which shows stellar yields specifically optimised for SN\,1987A, 
presents a much smaller mass (0.03\,\Msun) in the Si+S+Fe zone, which can reduce the SiS mass. 
However, only a factor of three difference in Si+S+Fe zone mass is insufficient to explain
the more than $10^3$ difference between predicted and estimated SiS mass.

We argue that a large SiO abundance and a small SiS abundance suggests the presence of some degrees 
of  mixing.
Figure\,\ref{fig-mixing} (b) show an example of a mixed case. 
This is expressed in one-dimension as a function of cumulative mass from the core to outer direction,
so that the clumpiness is not well reflected.
At a time of $10^7$\,s ($\sim$116 days) after the explosion,
Si and S atoms are  mixed with O atoms.
The mixing enables one to form SiO more efficiently from co-existing Si and O atoms 
Additionally, S can bond with O, forming SO and SO$_2$, rather than SiS in a mixed model \citep{Cherchneff:2009ex}.
Instead of mixing different zones completely, 
the mixing might have happened  only to an extent, with some clumps from one zone mixed with clumps from other zones.
Nevertheless, such mixing can still affect the chemical compositions, reducing  the SiS abundance and diverting Si to SiO.

An alternative possibility is using intrinsic Si to form SiO.  Some Si atoms are present throughout the stellar interior,
due to the Si incorporated into the star at the time of star-formation.
Chemical models \citep{Sarangi:2013bj}  predict that the abundance of SiO  to be $2\times10^{-8}$\,\Msun\, in O/C and He+O+C zones, where only intrinsic Si atoms are present
(Table 6 of their works). 
In these zones, Si atoms are mainly in silicate dust, instead of in SiO, and the predicted silicate dust mass 
is $4\times10^{-4}$\,\Msun. Opposed to the predictions of chemical model, 
if somehow the dust condensation rate of silicates can be lowered to a negligible level,
intrinsic Si can make an SiO mass as high as the measured mass.
However, that would  make it difficult to explain the large condensation rate of dust found in SN\,1987A  \citep{Matsuura:2011ij, Indebetouw:2013vv, Matsuura:2015kn}.
Further, while changing the dust condensation rate might increase the SiO abundance,  this does not solve the deficiency of SiS found in SN\,1987A.
Although there might be some uncertainty in the S chemistry (Cherchneff, private communication),
the presence of intrinsic Si and changing the dust condensation can potentially solve the problems of chemistry in SN\,1987A
partially, but not fully.

\subsection{Line profile and geometry}

The SiO line profiles show dips at the centre  (Fig.\ref{fig-lineprofile}).
The presence of dips requires a non-spherical distribution of SiO gas.
A filled sphere is unlikely to produce a dip in the line profile,
as even with optical depth effects it can only produce a flat top profile.
It might be possible to have a shell, an elongated ellipse, or possibly even a bipolar or torus shape.
The absence of a prominent dip in the CO line profile is compelling evidence that the CO and SiO have different spatial distributions.

Recently, \citet{Kjr:2010p29878} and \citet{Larsson:2013gx, Larsson:2016wg} investigated
the ejecta velocity structures, using the 1.64\,$\mu$m [Si\,{\small I}]+[Fe\,{\small II}] feature.
They found  asymmetric line profiles in both
[Si\,{\small I}]+[Fe\,{\small II}]  and H$\alpha$,
with  extents of over 3000\,km\,s$^{-1}$.
Such a wide profile is not found here for the SiO line.
In \citet{Kjr:2010p29878}, the red-shifted part of the line profile
is stronger than the blue-shifted component,
but such an extreme asymmetric structure is not found for SiO.
Moreover, a dip was not  found in either the [Si\,{\small I}]+[Fe\,{\small II}]  or the H$\alpha$ line,
\citep{Larsson:2016wg}.
Their line shapes do not resemble that of SiO.
The SiO line profile may arise from different parts of the ejecta from those traced by
 [Si\,{\small I}]+[Fe\,{\small II}]  and H$\alpha$ trace.
This is probably because line excitation mechanisms are different, as [Si\,{\small I}]+[Fe\,{\small II}] lines are produced by non-thermal excitation
from positrons, while SiO is most likely collisionally excited.

We argue that the SiO dip is associated with a fast wind triggered deep inside  the stellar core
at the time of the SN explosion.
Hydrodynamical simulations  by \citet{Hammer:2010di}
showed that an inhomogeneous distribution of entropy, triggered by shocks  within a second of the explosion \citep{Scheck:2008ja},
enhanced fast outflows of the inner core.
Elements from the inner zones, such as nickel and oxygen can overrun carbon from the helium envelope.
That shapes the ejecta into a multi-polar geometry.
Eventually, such turbulence ceases, so that the measured line velocities  a few months after the explosion are almost the same for all elements
\citep{McCray:1993p29839}.
Nevertheless, the asymmetric bipolar shape triggered immediately continues to be present in the spatial distributions of the gas.

\subsection{Isotope ratios} \label{section-isotope}

\begin{figure*}
\centering
\resizebox{0.7\hsize}{!}{\includegraphics{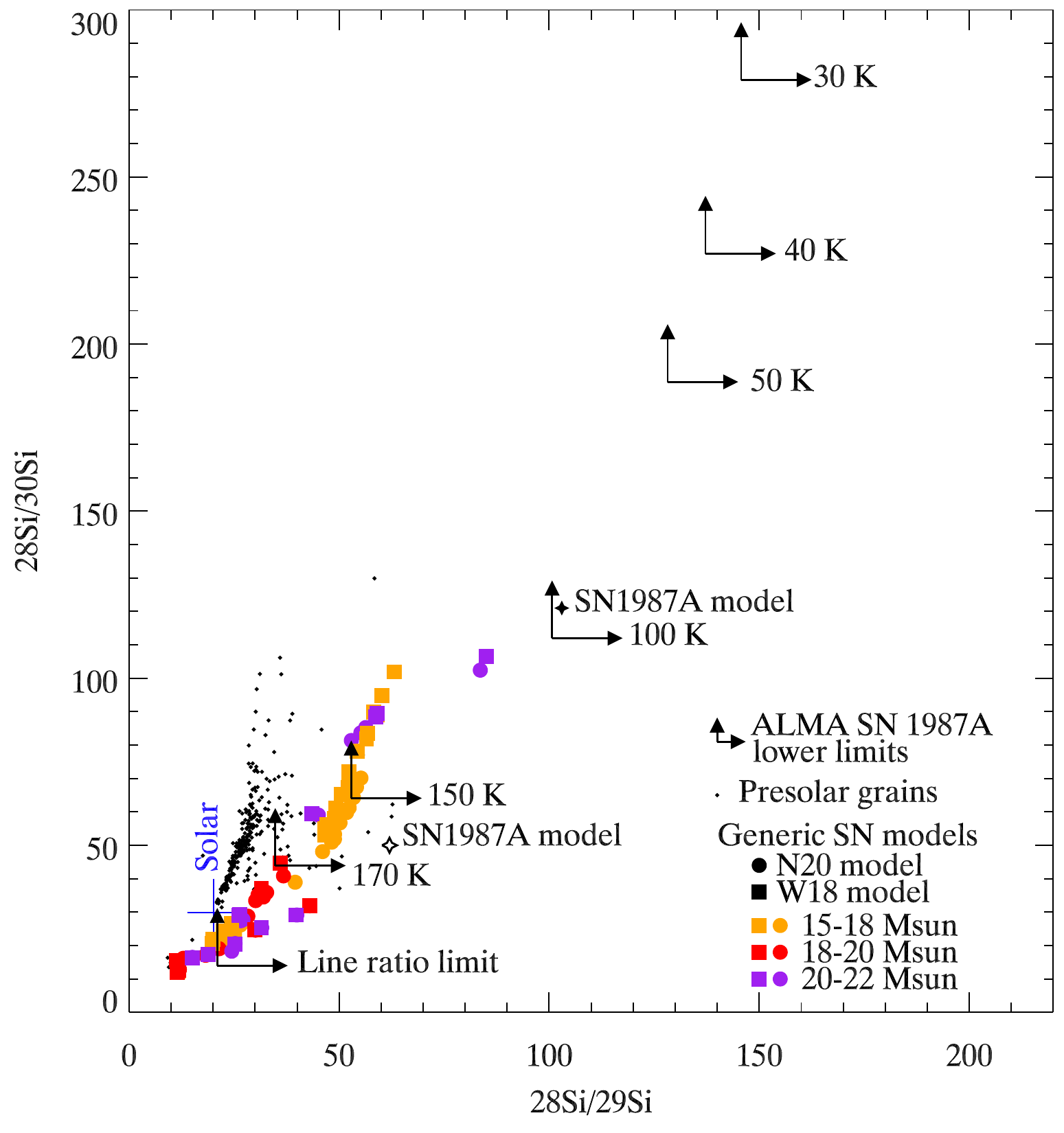}}
\caption{ 
The Si isotope ratios. ALMA lower limits for SN\,1987A are plotted with arrows, as a function of 
the assumed $T_{\rm {kin}}$.
The measured lower limits are compared with explosive nucleosynthesis models,
including SN\,1987A specific models (filled star: based on \citet{Woosley:1988p29819}; open star: based on \citet{Nomoto:2013js}).
Generic solar abundance models are from \citet{Sukhbold:2016bo},
as a function of zero-age main sequence mass. More details of these models are given in the main text.
Solar isotope ratios are indicated by blue lines.
The analysis of CO favours a temperature range of 30--50\,K. If such a low temperature
applies  to SiO as well, the Si isotope ratios of SN\,1987A could be rather a high value ($>128$).
The measured Si isotope ratios in pre-solar grains follow a different sequence from those
of the solar metallicity models.
\label{fig-isotope}}
\end{figure*}

Our ALMA spectrum covers the lines of SiO and CO isotopologues,
resulting in estimates of lower limits for the isotope ratios of SN\,1987A.

Together with the Crab Nebula, which has provided
direct measurements of  Ar isotope ratios with  the {\sc Herschel} Space Observatory
 \citep{Barlow:2013dr},
the remnant of SN\,1987A has allowed for the first time to estimate the isotope ratios
for core-collapse SNe \citep{Kamenetzky:2013fv}.

Explosive nucleosynthesis models have predicted the ratio of $^{28}$Si/$^{29}$Si for SN\,1987A.
\citet{Woosley:1995p29811} predicted  $^{28}$Si/$^{29}$Si to be 65,
while \citet{Thielemann:1996dw} predicted 8.
Our ALMA observations suggest $^{28}$Si/$^{29}$Si$>$13, which is larger than the ratio derived by 
 \citet{Thielemann:1996dw}.
More recent generic (i.e. not specific to SN\,1987A) SN models predict
$^{28}$Si/$^{29}$Si=62 and $^{28}$Si/$^{30}$Si=50 at  LMC metallicity \citep{Nomoto:2013js}.
These ratios  are much closer to the lower limits we extracted for SN 1987\,A with ALMA.


All the isotopes of Si are synthesised by oxygen burning, either before or during the explosion 
 \citep{Woosley:1988p29481, Woosley:1995p29811, Woosley:2002ck}.
The majority of $^{28}$Si is made by oxygen burning,
and adding neutrons results in the production of  $^{29}$Si and $^{30}$Si.
$^{29}$Si and $^{30}$Si can be synthesised further by neon burning. 
The ratio of $^{30}$Si/$^{28}$Si  depends on the 
neutron excess, which depends on the initial metallicity of the star \citep[e.g.][]{Woosley:1988p29481, Kobayashi:2011hja}. 
Lower metallicity gives a lower neutron excess,
resulting in a smaller production of neutron-rich isotopes.
Because $^{30}$Si requires more neutrons than $^{29}$Si, and even more than $^{28}$Si,
 $^{28}$Si/$^{30}$Si becomes even larger than $^{28}$Si/$^{29}$Si  at lower metallicity, compared to the solar metallicity.

Fig.\ref{fig-isotope} shows our ALMA lower limits for $^{28}$Si/$^{29}$Si and $^{28}$Si/$^{30}$Si, as a function
of the assumed {\sc RADEX} kinetic temperature.
These ALMA lower limits are compared with theoretically predicted ratios from explosive nucleosynthesis models
based on \citet{Sukhbold:2016bo}, with a more extensive reaction network included.
The range of initial stellar mass is limited to 15--22\,\Msun, close to the expected progenitor mass of SN\,1987A (18--20\,\Msun).
\citeauthor{Sukhbold:2016bo}'s model tested the explosion of supernovae at solar metallicity for 9--120\,\Msun\, stars, 
and their  models are calibrated against the stellar evolution models of SN\,1987A.
Here, only two types of models are plotted:
models calibrated against W18, which originated from \cite{Woosley:1988p29819}, 
 and N20 \citep{Saio:1988jq, Hashimoto:1989p29480, Shigeyama:1990es}.
All these models are for solar metallicity, as in \citet{Sukhbold:2016bo}.
We also plot the specific model for SN\,1987A, which is substantially sub-solar and based on the \citet{Woosley:1988p29481} model with up-to-date nuclear reaction rates with a more extensive reaction network.
The other SN\,1987A model is based on  \citet{Nomoto:2013js}.

Fig.\ref{fig-isotope} also includes measurements of the Si isotope ratios in pre-solar grains.
Pre-solar grains are dust grains that had been formed in stellar environments,
and are now found in primitive meteorites in the solar system \citep{Zinner:1998hd, 2014mcp..book..181Z}.
Some pre-solar grains  are indicative of SN origin.
The isotope ratios of pre-solar grains are taken from Washington University's data base \citep{Hynes:2009vp},
that assembled measurements from \citet{Hoppe:1994fn}, \citet{Zinner:1998hd}, 
\citet{Amari:1999eu} and \citet{Nittler:1996ci} etc.
The values plotted in Fig.\ref{fig-isotope} are for X-type SiC grains, whose types had been formed in supernova ejecta
\citep{Nittler:1996ci, Pignatari:2013dm}.

Our ALMA estimates of $^{28}$Si/$^{30}$Si  and $^{28}$Si/$^{29}$Si
lower limits have  SiO's $T_{\rm kin}$ dependence (Fig.\ref{fig-isotope}).
Although our SiO measurements have not well-constrained $T_{\rm kin}$, 
measurements and modelling of CO favour 20--50\,K.
If SiO indeed also has this temperature range,
the ALMA lower limits are $^{28}$Si/$^{29}$Si$>$128 and $^{28}$Si/$^{30}$Si$>$188, 
pointing to a  lower metallicity than  the solar metallicity models.
The limits are even below the SN\,1987A specific model, though the differences
are less than a factor of two.
Additionally, the SN\,1987A model, based on \citet{Sukhbold:2016bo}, used a different
semi-convection treatment from red-supergiant models. 
That could also contribute to a change of the oxygen-shell structure at the time of the explosion, potentially resulting in a different silicon isotope ratio. 
Subtle changes in the input parameters may potentially match the ALMA measurements.

If the isotope ratios of $^{28}$Si/$^{29}$Si and $^{28}$Si/$^{30}$Si are indeed $>$128 and $>$188, respectively,
they do not follow the sequence of these ratios in pre-solar grains.
This is reasonable, because neutron-rich isotopes are less efficiently produced at lower metallicity
\citep{Woosley:1995p29811}.

Obviously, our current interpretation of the ALMA results is limited
by the available SiO  isotopologue lines.
Ill-constrained excitation conditions are the main concern in this current work.
Detecting isotopologue lines and more SiO lines
would significantly improve the derived the abundance ratios.

Our ALMA measurements suggest carbon isotope ratios of $^{12}$C/$^{13}$C$>21$.
This lower limit  is consistent with the values predicted by explosive nucleosynthesis models.
The revised SN\,1987A models from \citet{Woosley:1988p29481} predict $^{12}$C/$^{13}$C=2482, whereas
\citet{Nomoto:2013js} predict  $^{12}$C/$^{13}$C=158 for 18\,\Msun\, and $Z=0.08$.
The solar value is 89.
It may be difficult to obtain a stronger constraint on $^{12}$C/$^{13}$C from observations, due to the very low abundance of $^{13}$C.

\section{Conclusions}

From ALMA spectral scans at 210--300\,GHz and 340--360\,GHz, we detected $^{12}$CO, $^{28}$SiO, HCO$^+$ and SO.
Upper limits for $^{29}$SiO and $^{30}$SiO, as well as $^{13}$CO, have been obtained.

The line profiles of SiO and CO clearly show differences, with a dip in the SiO line profile.
The presence of the dip clearly shows  that the SiO spatial distribution is asymmetric.

Our non-LTE models suggest the kinetic temperature and mass of CO to be 20--50\,K and 1.0--0.02\,\Msun,
depending on the temperature. The parameter range for SiO is less constrained, with a temperature range
of 20--170\,K and a mass range of $2\times10^{-3}$--$4\times10^{-5}$\,\Msun.
If the temperature of SiO and CO are more or less similar, the SiO mass range can be narrowed down to $2\times10^{-3}$--$5\times10^{-4}$\,\Msun.
With a 40\,K model, the upper limit for the HCO$^+$ mass is $5\times10^{-6}$\,\Msun,
and the SiS mass is $<6\times10^{-5}$\,\Msun. The HCO$^+$ mass is an upper limit because it has a potential line blending issue with SO$_2$,
although we expect that the dominant contributor is HCO$^+$

Together with an non-LTE model analysis and the detection of HCO$^+$, which is a tracer of dense molecular clouds,
the density of the line emitting region can be as high as 10$^5$--10$^6$\,cm$^{-3}$. 
Such a high density might be achieved, if the molecules are distributed in clumps with locally enhanced densities,
rather than being smoothly distributed across the ejecta.

Existing chemical models under-predict the HCO$^+$ mass, while the predicted SiS mass is far larger.
One possible explanation is that macroscopic mixing allowed some hydrogen clumps from the hydrogen-envelope to sink into
the C+O zone, and C and O clumps floated up into the hydrogen envelope.
That can enhance the probability of H$_2$ and CO reactions  forming HCO$^+$.
In the Si+S zone, some oxygen from the Si/O zone may have mixed in.

From our ALMA spectra, we have estimated lower limits for $^{28}$Si/$^{29}$Si and  $^{28}$Si/$^{30}$Si.
If the kinetic temperature of the SiO isotopologues are in the range of 20--50\,K, similar to CO,
its neutron rich isotopes appear to be less abundant in SN\,1987A,
consistent with the predictions of nucleosynthesis models at lower metallicity.


\section{acknowledgments}
We thank the referee for carefully reading the draft.
Prof. D. A. Williams and J.M.C. Rawlings are thanked for discussions.
MM and MJB acknowledge support from  UK STFC grant (ST/J001511/1), and MM is further supported
 by an STFC Ernest Rutherford fellowship (ST/L003597/1). 
 MJB acknowledges support  from European Research Council (ERC) Advanced Grant SNDUST 694520.
HLG acknowledges support from the European Research Council (ERC) in the form of Consolidator Grant {\sc CosmicDust} (ERC-2014-CoG-647939). 
SW acknowledges support  from  supported by NASA grant (NNX14AH34G).

\bibliography{sn1987a_alma}
\bibliographystyle{mn2e}

\label{lastpage}
\end{document}